\documentclass[12pt]{article}
\usepackage{amsfonts}
\usepackage{amsgen,amsmath,amstext,amsbsy,amsopn,amssymb}
\usepackage[dvips]{graphicx}
\usepackage{comment}
\usepackage{booktabs}
\usepackage[dvipsnames,table,dvipsnames*, svgnames*, hyperref]{xcolor}
\usepackage{epstopdf}
\usepackage{subfigure}

\newtheorem{Theorem}{Theorem}[section]
\newtheorem{Assumption}{Assumption}[section]

\newtheorem{Lemma}{Lemma}[section]
\newtheorem{Remark}{Remark}[section]

\newtheorem{Proposition}{Proposition}[section]

\newcommand{\be}{\begin{equation}}
\newcommand{\ee}{\end{equation}}

\newcommand{\beq}{\begin{equation}}
\newcommand{\eeq}{\end{equation}}
\newcommand{\beas}{\begin{eqnarray*}}
\newcommand{\eeas}{\end{eqnarray*}}
\newcommand{\bea}{\begin{eqnarray}}
\newcommand{\eea}{\end{eqnarray}}
\newcommand{\bei}{\begin{itemize}}
\newcommand{\eei}{\end{itemize}}
\newcommand{\ben}{\begin{enumerate}}
\newcommand{\een}{\end{enumerate}}
\newcommand{\bet}{\begin{theorem}}
\newcommand{\eet}{\end{theorem}}
\newcommand{\bel}{\begin{lemma}}
\newcommand{\eel}{\end{lemma}}
\newcommand{\bep}{\begin{proposition}}
\newcommand{\eep}{\end{proposition}}
\newcommand{\bed}{\begin{definition}}
\newcommand{\eed}{\end{definition}}
\newcommand{\bec}{\begin{corollary}}
\newcommand{\eec}{\end{corollary}}
\newcommand{\bex}{\begin{example}}
\newcommand{\eex}{\end{example}}

\newcommand {\RR}  {{\mathbb{R}}}

\renewcommand{\a}{{\mathbf{a}}}
\renewcommand{\b}{{\mathbf{b}}}

\renewcommand{\u}{{\mathbf{u}}}
\renewcommand{\v}{{\mathbf{v}}}

\newcommand{\x}{{\mathbf{x}}}
\newcommand{\y}{{\mathbf{y}}}

\newcommand{\A}{{\mathbf{A}}}
\newcommand{\B}{{\mathbf{B}}}
\newcommand{\C}{{\mathbf{C}}}
\newcommand{\D}{{\mathbf{D}}}

\newcommand{\I}{{\mathbf{I}}}

\newcommand{\Q}{{\mathbf{Q}}}

\renewcommand{\S}{{\mathbf{S}}}
\newcommand{\U}{{\mathbf{U}}}
\newcommand{\V}{{\mathbf{V}}}

\newcommand{\var}{{\rm var}}

\newcommand{\X}{{\mathbf{X}}}

\newcommand{\Z}{{\mathbf{Z}}}
\newcommand{\SSigma}{{\mathbf{\Sigma}}}

\newcommand{\bbeta}{\boldsymbol{\beta}}

\newcommand{\GGamma}{\boldsymbol{\Gamma}}
\newcommand{\m}{\boldsymbol{\mu}}

\newcommand{\bDelta}{\boldsymbol{\Delta}}

\def\trans{^{\top}}

\newcommand{\argmin}{\mathop{\rm arg\min}}

\newcommand{\tr}{{\rm tr}}
\newcommand{\supp}{{\rm supp}}
\newcommand{\Ex}{{\rm E}}

\newcommand{\diag}{{\rm diag}}

\begin{document}

\begin{title}
{Minimax Rate-optimal Estimation of High-dimensional  Covariance Matrices with Incomplete Data
		\footnote{T. Tony Cai is Dorothy Silberberg Professor of Statistics, Department of Statistics, The Wharton School, University of Pennsylvania, Philadelphia, PA (E-mail: tcai@wharton.upenn.edu); Anru Zhang is Assistant Professor, University of Wisconsin-Madison, Madison, WI (E-mail: anruzhang@stat.wisc.edu). The research of Tony Cai and Anru Zhang  was supported in part by NSF Grant DMS-1208982 and DMS-1403708, and NIH Grant R01 CA127334.}
}
\end{title}

\author{T. Tony Cai \, and \, Anru Zhang\\
} 
\date{} 
\maketitle

	\begin{abstract}
		Missing data occur frequently in a wide range of applications. In this paper, we consider estimation of high-dimensional  covariance matrices in the presence of missing observations under a general missing completely at random model in the sense that the missingness is not dependent on the values of the data. Based on incomplete data, estimators for bandable and sparse covariance matrices  are proposed and their theoretical and numerical properties are investigated.  
		
		Minimax rates of convergence are established under the spectral norm loss and the proposed estimators are shown to be rate-optimal under mild regularity conditions. Simulation studies demonstrate that the estimators perform well numerically.  The methods are also illustrated through an application to data from four ovarian cancer studies. The key technical tools developed in this paper are of independent interest and potentially useful for a range of related problems in high-dimensional statistical inference with missing data.

	\end{abstract}
	
	\noindent{\it Keywords:\/}
	Adaptive thresholding,  bandable covariance matrix,  generalized sample covariance matrix, missing data, optimal rate of convergence, sparse covariance matrix, thresholding.

	\section{Introduction}
	
	The problem of missing data arises frequently in a wide range of  fields, including biomedical studies, social science, engineering, economics,  and computer science. Statistical inference in the presence of missing observations has been well studied in classical statistics. See, e.g., Ibrahim and Molenberghs \cite{Ibrahim} for a review of missing data methods in longitudinal studies and Schafer \cite{Schafer} for  literature on handling multivariate data with missing observations. See Little and Rubin \cite{book_misssing_value} and the references therein for a comprehensive treatment of missing data problems.
	
	Missing data also occurs in contemporary high-dimensional inference problems, whose dimension $p$ can be comparable to or even much larger than the sample size $n$.  For example, in large-scale genome-wide association studies (GWAS), it is common for many subjects to have missing values on some genetic markers due to various reasons,  including insufficient resolution, image corruption, and experimental error during the laboratory process.
	Also, different studies may have different volumes of genomic data available by design. For instance, the four genomic ovarian cancer studies discussed in Section 4 have throughput measurements of mRNA gene expression levels, but only one of these also has microRNA measurements (Cancer Genome Atlas Research Network \cite{cancer2011integrated}, Bonome et al. \cite{bonome2005expression}, Tothill et al. \cite{tothill2008novel} and Dressman et al. \cite{dressman2007integrated}). 
	Discarding samples with any missingness is  highly inefficient and could induce bias due to non-random missingness. 
	It is of significant interest to integrate multiple high-throughput studies of the same disease, not only to boost statistical power but also to improve the biological interpretability. However, considerable challenges arise when integrating such studies due to missing data.
	
	
	Although there have been significant recent efforts to develop methodologies and theories for high dimensional data analysis, there is a paucity of methods with theoretical guarantees for statistical inference with missing data in the high-dimensional setting. Under the assumption that the components are missing uniformly  and completely at random (\emph{MUCR}), Loh and Wainwright \cite{Loh} proposed a non-convex optimization approach to high-dimensional linear regression, Lounici \cite{Lounici2} introduced a method for estimating a low-rank covariance matrix and  Lounici \cite{Lounici1} considered  sparse principal component analysis. In these papers, theoretical properties of the procedures were analyzed. These methods and theoretical results critically depend on the MUCR assumption. 
	
	Covariance structures play a fundamental role in 
	high-dimensional statistics. It is of direct interest in a wide range of applications including genomic data analysis, particularly for hypothesis generation.  Knowledge of the covariance structure is critical to many statistical methods, including discriminant analysis, principal component analysis, clustering analysis, and regression analysis. In the high-dimensional setting with complete data, inference on the covariance structure has been actively studied in recent years. See Cai, Ren and Zhou \cite{Cai_Ren_Zhou_survey} for a survey of recent results on minimax and adaptive estimation of high-dimensional covariance  and precision matrices under various structural assumptions. 
	Estimation of high-dimensional covariance matrices in the presence of missing data also has wide applications in biomedical studies, particularly in integrative genomic analysis which holds great potential in providing a global view of genome function (see Hawkins et al. \cite{Hawkins}).  
	
	In this paper, we consider estimation of high-dimensional covariance matrices in the presence of missing observations under a general missing completely at random (\emph{MCR}) model in the sense that the missingness is not dependent on the values of the data. 
	Let  $\X_1, \ldots, \X_{n}$ be $n$ independent copies of a $p$ dimensional random vector $\X$ with mean $\m$ and
	covariance matrix $\SSigma$. 
	Instead of observing the complete sample $\{\X_1, \ldots, \X_{n}\}$, one observes the sample with missing values, where the observed coordinates of $\X_k$ are indicated by a vector $\S_k\in  \{0, 1\}^p$, $k=1, ..., n$. That is,
	\begin{equation}\label{eq:missing_S}
		X_{jk} \text{ is observed if } S_{jk}=1 \;\mbox{ and } \;  X_{jk} \text{ is missing if } S_{jk}=0.
	\end{equation} 
	Here $X_{jk}$ and $S_{jk}$ are respectively the $j$th coordinate of the vectors $\X_k$ and $\S_k$.
	We denote the incomplete sample with missing values by  $\X^* = \{\X_1^*, \ldots,   \X_n^*\}$. 
	The major goal of the present paper is to estimate $\SSigma$, the covariance matrix of  $\X$, with theoretical guarantees  based on the incomplete data $\X^*$ in the high-dimensional setting where $p$ can be much larger than $n$.

	This paper focuses on estimation of  high-dimensional bandable covariance matrices and sparse covariance matrices in the presence of missing data. These two classes of covariance matrices arise frequently in many applications, including genomics, econometrics, signal processing, temporal and spatial data analyses, and chemometrics.
	Estimation of these high-dimensional structured covariance matrices have been well studied in  the setting of complete data in a number of recent papers, e.g.,  Bickel and Levina \cite{Bickel,Bickel2}, Karoui \cite{ElKaroui_sparse_cov}, Rothman et al. \cite{Rothman_sparse_cov}, Cai and Zhou \cite{Cai_Zhou_covariance}, Cai and Liu \cite{Cai_Liu}, Cai et al. \cite{Cai_Ma_Wu2,Cai_Zhang_Zhou_bandable} and Cai and Yuan \cite{Cai_Yuan}.  Given an incomplete sample $\X^* $ with missing values, we introduced a ``generalized" sample covariance matrix, which can be viewed as an analog of the usual sample covariance matrix in the case of complete data. For estimation of bandable covariance matrices, where the entries of the matrix decay as they move away from the diagonal, a blockwise tridiagonal estimator is introduced and is shown to be rate-optimal. We then consider estimation of sparse covariance matrices. An adaptive thresholding estimator based on the generalized sample covariance matrix is proposed. The estimator is shown to achieve the optimal rate of convergence over a large class of approximately sparse covariance matrices under mild conditions.
	
	The technical analysis for the case of missing data is much more challenging than that for the complete data, although some of the basic ideas are similar. To facilitate the theoretical analysis of the proposed estimators, we establish two key technical results, first, a large deviation result for a sub-matrix of the generalized sample covariance matrix and second, a large deviation bound for the self-normalized entries of the generalized sample covariance matrix. These technical tools are not only important for the present paper, but also useful for other related problems in high-dimensional statistical inference with missing data.
	

	A simulation study is carried out to examine the numerical performance of  the proposed estimation procedures. The results show that the proposed estimators perform well numerically. Even in the MUCR setting, our proposed procedures for estimating bandable, sparse covariance matrices, which do not rely on the information of the missingness mechanism,  outperform the ones specifically  designed for MUCR. The advantages are more significant under the setting of missing completely at random but not uniformly.
	We also illustrate our procedure with an application to data from four ovarian cancer studies that have different volumes of genomic data by design. The proposed estimators enable us to estimate the covariance matrix by integrating the data from all four studies and lead to a more accurate estimator. Such high-dimensional covariance matrix estimation with missing data is also useful for other types of data integration. See further discussions in Section \ref{real.data.sec}.
	
	The rest of the paper is organized as follows. Section \ref{bandable-spiked.covariance.sec} considers estimation of bandable covariance matrices with incomplete data. The minimax rate of convergence is established for the spectral norm loss under regularity conditions. Section \ref{sparse.covariance.sec} focuses on estimation of high-dimensional sparse covariance matrices and introduces an adaptive thresholding estimator in the presence of missing observations. Asymptotic properties of the estimator under the spectral norm loss is also studied. 
	Numerical performance of the proposed methods is investigated in Section \ref{simulation.sec} through both simulation studies and an analysis of an ovarian cancer dataset. Section \ref{discussion.sec} discusses a few related problems. Finally the proofs of the main results are given in Section \ref{proof.sec} and the Supplement.
	
		\section{Estimation of Bandable Covariance Matrices}
		\label{bandable-spiked.covariance.sec}
		In this section, we consider estimation of  bandable covariance matrices with incomplete data. Bandable covariance matrices, whose entries decay as they move away from the diagonal, arise frequently in temporal and spatial data analysis. See, e.g., Bickel and Levina \cite{Bickel} and Cai et al. \cite{Cai_Ren_Zhou_survey} and the references therein. 
		The procedure relies on a ``generalized" sample covariance matrix. We begin with basic notation and definitions that will be used throughout the rest of the paper.
		
		\subsection{Notation and Definitions}\label{sec.notation}

		Matrices and vectors are denoted by boldface letters. For a vector $\bbeta \in \RR^p$, we denote the Euclidean $q$-norm by $\|\bbeta\|_q$, i.e., $\|\bbeta\|_q = \sqrt[q]{\sum_{i=1}^p |\beta_i|^q}$. 
		Let $\A = \U\D \V\trans = \sum_i \lambda_i(\A)\u_i\v_i\trans$ be the singular value decomposition of a matrix $\A\in \RR^{p_1\times p_2}$,  where $\D= \diag\{\lambda_1(\A), \ldots\}$ with $\lambda_1(\A) \geq \cdots\ge 0$ being the singular values. For $1\leq q\leq \infty$, the Schatten-$q$ norm $\|\A\|_q$ is defined by $\|\A\|_q = \{\sum_i \lambda_i^q(\A)\}^{1/q}$. In particular, $\|\A\|_2 = \sqrt{\sum_i \lambda_i^2(\A)}$ is the Frobenius norm of $\A$ and will be denoted as $\|\A\|_F$; $\|\A\|_\infty = \lambda_1(\A)$ is the spectral norm of $\A$ and will be simply denoted as $\|\A\|$. For $1\leq q\leq \infty$ and $A\in \mathbb{R}^{p_1\times p_2}$, we denote the operator $\ell_q$ norm of $\A$ by $\|\A\|_{\ell_q}$ which is defined as $\|\A\|_{\ell_q} = \max_{x\in \RR^{p_2}}\|\A\x\|_q/\|\x\|_q$. The following are well known facts about the various norms of a matrix $A=(a_{ij})$, 
		\begin{equation}
		\|\A\|_{\ell_1} = \max_j \sum_{i=1}^{p_1} |a_{ij}|,\quad \|\A\|_{\ell_2} = \|\A\| = \lambda_1(\A),\quad \|\A\|_{\ell_\infty} = \max_i \sum_{j=1}^{p_2} |a_{ij}|,
		\end{equation}
		and, if $\A$ is symmetric, $\|\A\|_{\ell_1} = \|\A\|_{\ell_\infty} \geq \|\A\|_{\ell_2}$. When $R_1$, $R_2$ are two subsets of $\{1,\ldots, p_1\}$, $\{1,\ldots, p_2\}$ respectively, we note $\A_{R_1\times R_2} = (a_{ij})_{i\in R_1, j \in R_2}$ as the sub-matrix of $\A$ with indices $R_1$ and $R_2$. In addition, we simply write $\A_{R_1\times R_1}$ as $\A_{R_1}$.
		
		We denote by $\X_1, \ldots, \X_n$ a complete random sample (without missing observations) from a $p$-dimensional distribution with mean $\m$ and covariance matrix $\SSigma$. The sample mean and sample covariance matrix are defined as
		\begin{equation}
		\bar \X = \frac{1}{n}\sum_{k=1}^n \X_k, \quad \hat \SSigma =  \frac{1}{n}\sum_{k=1}^n \left(\X_{k} - \bar \X\right)\left(\X_{k} - \bar \X\right)\trans.
		\end{equation}

		Now we introduce the notation related to the incomplete data with missing observations. Generally, we use the superscript ``$\ast$" to denote objects related to missing values. Let $\S_1, ..., \S_n$ be the indicator vectors for the observed values  (see \eqref{eq:missing_S}) and let $\X^* = \{\X_1^*, \ldots,  \X_n^*\}$ be the observed incomplete data where the observed entries are indexed by the vectors $\S_1, ..., \S_n\in \{0, 1\}^p$.
		In addition, we define
		\begin{equation}
		n_{ij}^* = \sum_{k=1}^n S_{ik}S_{jk}, \quad 1\leq i, j\leq p.
		\end{equation} 
		Here $n_{ij}^*$ is the number of  vectors $\X_k^*$ in which the $i^{th}$ and $j^{th}$ entries are both observed. 
		For convenience, we also denote 
		\begin{equation}
		n_i^* = n_{ii}^*, \quad n_{\min}^* = \min_{i, j} n_{ij}^*.
		\end{equation}
		
		Given a sample $\X^* = \{\X_1^*, \ldots,   \X_n^*\}$ with missing values,  the sample mean and sample covariance matrix can no longer be calculated in the usual way. Instead, we propose the ``generalized sample mean" $\bar \X^*$ defined by
		\begin{equation}\label{eq:bar_X}
		\bar \X^* = (\bar X_{i}^*)_{1\leq i\leq p} \quad\mbox{with}\quad  \bar X_{i}^* = \frac{1}{n_{i}^*}\sum_{k=1}^n X_{ik}S_{ik}, \quad 1\leq i\leq p,
		\end{equation}
		where $X_{ik}$ is the $i$th entry of $\X_k$,
		and the ``generalized sample covariance matrix" $\hat\SSigma^*$ defined by
		\begin{equation}\label{eq:hat_Sigma}
		\hat\SSigma^* = (\hat\sigma_{ij}^*)_{1\leq i,j\leq p} \quad\mbox{with}\quad \hat \sigma_{ij}^* = {1\over n_{ij}^*}\sum_{k=1}^n(X_{ik} - \bar X_{i}^*)(X_{jk} - \bar X_{j}^*)S_{ik}S_{jk}.
		\end{equation}

		As will be seen later, the generalized sample mean $\bar \X^*$ and the generalized sample covariance matrix $\hat\SSigma^*$ play  similar roles as those of the conventional sample mean and sample covariance matrix in inference problems, but the technical analysis can be much more involved. 
		Some distinctions between the generalized sample covariance matrix $\hat\SSigma^*$ and the usual sample covariance matrix $\hat\SSigma$ are that $\hat\SSigma^*$ is in general not non-negative definite, and each entry $\hat \sigma_{ij}^\ast$ is the average of a varying number ($n_{ij}^\ast$) of samples, which create additional  difficulties in the technical analysis.
		
		
		Regarding the mechanism of missingness, the assumption we use for the theoretical analysis is \emph{missing completely at random}. This is a more general setting than the one considered previously by Loh and Wainwright \cite{Loh} and Lounici \cite{Lounici1}.
		\begin{Assumption}[Missing Completely at Random (MCR)]\label{as:MCR} \rm
			$\S = \{\S_1, \ldots, \S_n\}$ is not dependent on the values of $\X$. Here $\S$ can be either deterministic or random, but independent of $\X$.
		\end{Assumption}
		
		We adopt Assumption 1 in Chen et al. \cite{Chen} and assume that the random vector $\X$ is sub-Gaussian satisfying the following assumption.
		\begin{Assumption}[Sub-Gaussian Assumption]\label{as:sub-Gaussian} \rm
			$\X = \{\X_1, \ldots, \X_n\}$. Here the columns $\X_k$ are i.i.d. and can be expressed as
			\begin{equation}
			\label{SubGaussian-Z}
			\X_k = \GGamma \Z_k + \m, \quad k = 1,\ldots, n,
			\end{equation}
			where $\m$ is a fixed $p$-dimensional mean vector, $\GGamma \in \mathbb{R}^{p\times q}$ is a fixed matrix with $q\geq p$ so that $\GGamma\GGamma\trans = \SSigma$, $\Z_k = (Z_{1k}, \ldots, Z_{mk})\trans$ is an $m$-dimensional random vector with the components mean 0, variance 1, and i.i.d. sub-Gaussian, with the exception of i.i.d. Rademacher. More specifically, each $Z_{ik}$ satisfies that $\Ex Z_{ik}=0, \var(Z_{ik}) = 1$, $0<\var(Z_{ik}^2) <\infty$, and there exists $\tau>0$ such that $\Ex e^{t Z_{ik}} \leq \exp(\tau t^2/2)$ for all $t>0$. 
		\end{Assumption}
		Note that the exclusion of the Rademacher distribution in Assumption \ref{as:sub-Gaussian} is only required for estimation of sparse covariance matrices. See Remark \ref{Rademacher.rmk} for further discussions.
		
		\subsection{Rate-optimal Blockwise Tridiagonal Estimator}
		
		We  follow Bickel \cite{Bickel} and Cai et al. \cite{Cai_Zhang_Zhou_bandable} and consider estimating the covariance matrix $\SSigma$ over the parameter space $\mathcal{U}_\alpha = \mathcal{U}_\alpha(M_0, M) $ where
		\begin{equation}
		\mathcal{U}_\alpha(M_0, M) = \left\{\SSigma: \max_j \sum_i\{|\sigma_{ij}|: |i-j| > k\} \leq Mk^{-\alpha} \text{ for all } k, \|\SSigma\|\leq M_0 \right\}.
		\end{equation}
		
\begin{figure}[htbp]
			\centerline{\includegraphics[scale=0.43]{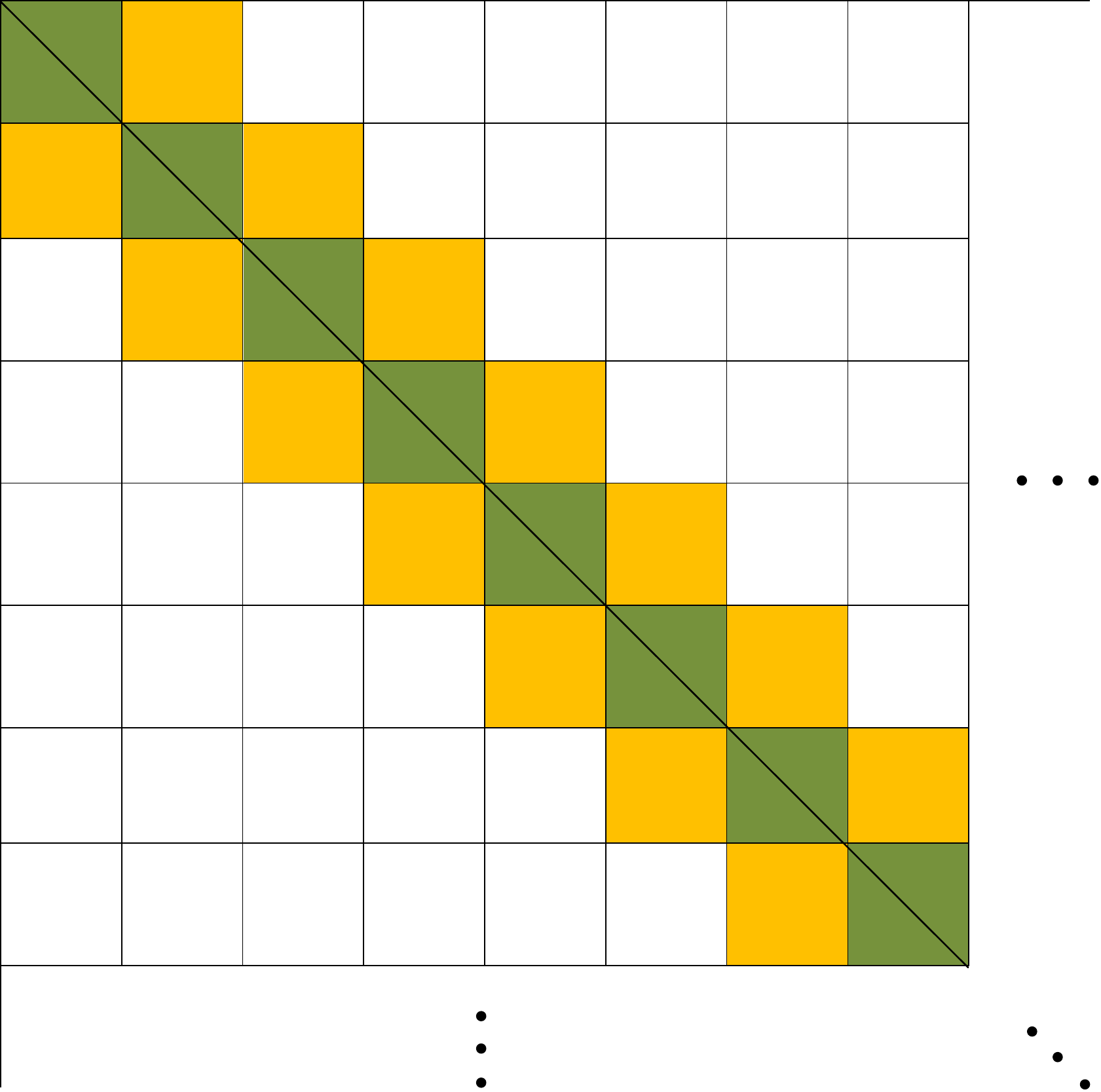}}
			\caption{Weight matrix for the blockwise tridiagonal  estimator.}
			\label{fig:tridiag}
		\end{figure}

		Suppose we have $n$ i.i.d. samples with missing values $\X_1^\ast, \ldots, \X_n^\ast$ with covariance matrix $\SSigma\in \mathcal{U}_\alpha(M_0, M)$. 
		We propose a blockwise tridiagonal estimator $\hat{\SSigma}^{\rm bt}$ to estimate $\SSigma$. We begin by dividing the generalized sample covariance matrix $\hat\SSigma^\ast$ given by \eqref{eq:hat_Sigma} into blocks of size $k\times k$ for some $k$. More specifically, pick an integer $k$ and let $N = \lceil p/k\rceil$. Set $I_j = \{(j-1)k+1, \ldots, jk\}$ for $1\leq j\leq N-1$, and $I_N = \{(N-1)k+1, \ldots, p\}$. For $1\leq j, j' \leq N$ and $\A = (a_{i_1, i_2})_{p\times p}$, define
		$$\A_{I_j\times I_{j'}} = (a_{i_1, i_2})_{i_1\in I_j, i_2\in I_{j'}} $$
		and define the blockwise tridiagonal estimator $\hat{\SSigma}^{\rm bt}$ by
		\begin{equation}
		\hat{\SSigma}_{I_j\times I_{j'}} = \left\{\begin{array}{ll}
		\hat\SSigma^\ast_{I_j\times I_{j'}}, & \text {if  } |j - j'| \leq 1;\\
		0, & \text{otherwise. }
		\end{array}
		\right.
		\end{equation}
		That is, $\hat{\SSigma}_{I_j \times I_{j'}}$ is estimated by its sample counterpart if and only if $j$ and $j'$ differ by at most 1.
		The weight matrix of the blockwise tridiagonal estimator $\hat{\SSigma}^{\rm bt}$ is illustrated in Figure \ref{fig:tridiag}.

		\begin{Theorem}\label{th:bandable}
			Suppose Assumptions \ref{as:MCR} and  \ref{as:sub-Gaussian} hold.
			Then, conditioning on $\S$,  the blockwise tridiagonal $\hat \SSigma^{\rm bt}$ with $k = (n_{\min}^\ast)^{1/(2\alpha+1)}$ satisfies
			\begin{equation}
			\label{BT.rate}
			\sup_{\SSigma \in \mathcal{U}_\alpha(M, M_0)} \Ex\|\hat{\SSigma}^{\rm bt} - \SSigma\|^2 \leq C(n_{\min}^\ast)^{-2\alpha/(2\alpha+1)} + C\frac{\ln p}{n_{\min}^\ast},
			\end{equation}
			where $C$ is a constant depending only on $M$, $M_0$, and $\tau$ from Assumption \ref{as:sub-Gaussian}. 
		\end{Theorem}
				The optimal choice of block size $k$ depends on the unknown ``smoothness parameter" $\alpha$. In practice, $k$ can be chosen by cross-validation. See Section \ref{CV.sec} for further discussions. Moreover, the convergence rate in \eqref{BT.rate} is optimal as we also have the following lower bound result.
		\begin{Proposition}\label{pr:bandable_lower} For any $n_0\geq 1$ such that $p\leq \exp(\gamma n_0)$ for some constant $\gamma>0$, conditioning on $\S$ we have
		$$ \inf_{\hat{\SSigma}}\sup_{\substack{\SSigma \in \mathcal{U}_{\alpha}(M, M_0)\\ \S: n_{\min}^\ast \geq n_0}} \Ex \left(\|\hat{\SSigma} - \SSigma\|^2\right) \geq C(n_0)^{-2\alpha/(2\alpha+1)} + C\frac{\ln p}{n_0}. $$ 
		\end{Proposition}
		

		\begin{Remark}[Tapering and banding estimators] {\rm 
				It should be noted that the same rate of convergence can also be attained by tapering and banding estimators with  suitable choices of tapering and banding parameters. Specifically,
				let $\hat \SSigma^{\rm tp}$ and $\hat \SSigma^{\rm bd}$ be respectively the tapering and banded estimators proposed in Cai et al. \cite{Cai_Zhang_Zhou_bandable} and Bickel and Levina \cite{Bickel} with
				\begin{equation}
				\hat \SSigma^{\rm tp} = \hat \SSigma^{\rm tp}_k = (w_{ij}^{\rm tp} \hat\sigma_{ij}^\ast)_{1\leq i, j\leq p} \quad \text{and} \quad \hat \SSigma^{\rm bd} = \hat \SSigma^{\rm bd}_k = (w_{ij}^{\rm bd} \hat{\sigma}_{ij}^\ast)_{1\leq i, j \leq p},
				\end{equation}
				where $w^{\rm tp}_{ij}$ and $w^{\rm bd}_{ij}$ are the weights defined as
				\begin{equation}
				w_{ij}^{\rm tp} = \left\{\begin{array}{ll}
				1, & \text{when } |i-j|  \leq k/2,\\
				2 - \frac{|i-j|}{k_h}, & \text{when } k/2< |i-j| < k\\
				0, & \text{otherwise}
				\end{array}
				\right. \quad \text{and} \quad
				w_{ij}^{\rm bd} = \left\{\begin{array}{ll}
				1, & \text{when } |i-j|  \leq k,\\
				0, & \text{otherwise}
				\end{array}
				\right.. 
				\end{equation}
				Then the estimators $\hat{\SSigma}^{\rm tp}$ and $\hat{\SSigma}^{\rm bd}$ with $k = (n_{\min}^\ast)^{1/(2\alpha + 1)}$ attains the  rate given in \eqref{BT.rate}.
			}
		\end{Remark}

The proof of Theorem \ref{th:bandable} shares some basic ideas with that for the complete data case (See, e.g. Theorem 2 in Cai et al. \cite{Cai_Zhang_Zhou_bandable}). However, it relies on a  new key technical tool which is a large deviation result for a sub-matrix of the generalized sample covariance matrix under the spectral norm. This random matrix result for the case of missing data, stated in the following lemma,  can be potentially useful for other, related high-dimensional missing data problems.
The proof of Lemma \ref{lm:covariance_spectral_b}, given in Section \ref{proof.sec}, is more involved than the complete data case, as in the generalized sample covariance matrix each entry, $\hat\sigma_{ij}^\ast$, is the average of a varying number of samples.

		\begin{Lemma}\label{lm:covariance_spectral_b}
			Suppose Assumptions \ref{as:MCR} and  \ref{as:sub-Gaussian} hold.
			Let $\hat \SSigma^*$ be the generalized sample covariance matrix defined in \eqref{eq:hat_Sigma} and let $A$ and $B$ be two subsets of $\{1, \ldots, p\}$. Then, conditioning on $\S$, the submatrix $\hat \SSigma^*_{A\times B}$ satisfies
			\begin{equation}
			\begin{split}
			& \Pr\left(\|\hat\SSigma^\ast_{A \times B} - \SSigma_{A \times B}\| \leq x\right)\\
			\geq & 1- C\cdot (49)^{|A\cup B|}\exp\left\{-cn_{\min}^\ast\min\left(\frac{x^2}{\tau^4\|\SSigma_{A}\|\|\SSigma_B\|}, \frac{x}{\tau^2\left(\|\SSigma_{A}\|\|\SSigma_B\|\right)^{1/2}}\right)\right\}
			\end{split}
			\end{equation}
			for all $x>0$. Here $C>0$ and $c>0$ are two absolute  constants.
		\end{Lemma}

	\section{Estimation of Sparse Covariance Matrices}
	\label{sparse.covariance.sec}
	
	In this section, we consider  estimation of high-dimensional sparse covariance matrices in the presence of missing data. We introduce an adaptive thresholding estimator based on incomplete data and investigate its asymptotic properties.

	\subsection{Adaptive Thresholding Procedure}
	
	Sparse covariance matrices arise naturally in a range of applications including genomics. Estimation of sparse covariance matrices has been considered in several recent papers in the setting of complete data (see, e.g., Bickel and Levina \cite{Bickel2}, El Karoui \cite{ElKaroui_sparse_cov}, Rothman et al. \cite{Rothman_sparse_cov}, Cai and Zhou \cite{Cai_Zhou_covariance} and Cai and Liu \cite{Cai_Liu}).  Estimation of a sparse covariance matrix is intrinsically a heteroscedastic problem in the sense that the variances of the entries of the sample covariance matrix can vary over a wide range.  To treat the heteroscedasticity of the sample covariances, Cai and Liu \cite{Cai_Liu} introduced an adaptive thresholding procedure which adapts to the variability of the individual entries of the sample covariance matrix and outperforms the universal thresholding method. The estimator is shown to be simultaneously rate optimal over collections of sparse covariance matrices.
	
	In the present setting of missing data, the usual sample covariance matrix is not available. Instead we apply the idea of adaptive thresholding to the generalized sample covariance matrix $\hat \SSigma^*$.
	The procedure can be described as follows. Note that $\hat \SSigma^*$ defined in \eqref{eq:hat_Sigma} is a nearly unbiased estimate of $\SSigma$, we may write it element-wise as 
	\[
	\hat \sigma_{ij}^* \approx \sigma_{ij} + \sqrt{\frac{\theta_{ij}}{n_{ij}^\ast}}z_{ij}, \quad 1 \le i, j \le p,
	\]
	where $z_i$ is approximately normal with mean 0 and variance 1, and $\theta_{ij}$ describes the uncertainty of estimator $\sigma_{ij}^\ast$ to $\sigma_{ij}$ such that
	$$\theta_{ij} = \var\left\{(X_i - \mu_i)(X_j - \mu_j) - \sigma_{ij}\right\}.$$ 
	We can estimate $\theta_{ij}$ by
	\begin{equation}\label{eq:hat_theta}
		\hat \theta_{ij}^* = \frac{1}{n_{ij}^*}\sum_{k = 1}^n \left\{(X_{ik} - \bar X_{i}^*)(X_{jk} - \bar X_{j}^*) - \hat\sigma_{ij}^*\right\}^2S_{ik}S_{jk}.
	\end{equation}
	Lemma \ref{lm:covariance} given at the end of this section shows that  $\hat \theta_{ij}^*$ is a good estimate of   $\theta_{ij}$.
	
	Since the covariance matrix $\SSigma$ is assumed to be sparse, it is natural to estimate $\SSigma$ by individually thresholding
	$\hat \theta_{ij}^*$ according to its own variability as measured by $\hat \theta_{ij}^*$.
	Define the thresholding level $\lambda_{ij}$ by
	$$\lambda_{ij} = \delta\sqrt{\frac{\hat \theta_{ij}^*\ln p}{n_{ij}^*}}, \quad 1\leq i,j\leq p,$$
	where $\delta$ is a thresholding constant which can be taken as  2.

	Let $T_\lambda$ be a thresholding function satisfying the following conditions, 
	\begin{enumerate}
		\item[(1).] $|T_\lambda(z)|\leq c_T|y|$ for all $z,y$ such that $|z-y|\leq \lambda$;
		\item[(2).]  $T_\lambda(z) = 0$ for $|z|\leq \lambda$;
		\item[(3).]  $|T_\lambda(z)-z|\leq \lambda$, for all $z\in \RR$.
	\end{enumerate}
	These conditions are met by many well-used thresholding functions, including the soft thresholding rule $T_\lambda(z)={\rm sgn}(z)(z-\lambda)_{+}$, where sgn$(z)$ is the sign function such that sgn$(z)=1$ if $z > 0$, sgn$(z) = 0$ if $z=0$, and sgn$(z) = -1$ if $z < 0$, 
	and the adaptive lasso rule $T_\lambda(z)=z(1-|\lambda/z|^{\eta})_+$ with
	$\eta\geq 1$ (see Rothman et al. \cite{Rothman_sparse_cov}). 
	The hard thresholding function does not satisfy Condition (1), but our analysis also applies to hard thresholding under similar conditions. 
	
	The covariance matrix $\SSigma$ is estimated by $\hat\SSigma^{\rm at} = (\hat{\sigma}^{\rm at}_{ij})_{1\leq i,j\leq p}$ where $\hat{\sigma}^{\rm at}_{ij}$ is the thresholding estimator defined by
	\begin{equation}
		\hat\sigma^{\rm at}_{ij} = T_{\lambda_{ij}}(\hat{\sigma}_{ij}^*).
	\end{equation}
	Note that here each entry $\hat{\sigma}_{ij}^*$ is thresholded according to its own variability.
	
	\subsection{Asymptotic Properties}
	
	We now investigate the properties of the thresholding estimator $\hat\SSigma^{\rm at}$ over the following parameter space for sparse covariance matrices,
	\begin{equation}\label{eq:Hc_np}
		\mathcal{H}(c_{n,p})=\left\{ \SSigma=(\sigma_{ij}) :\max_{1\leq i\leq p}\sum_{j=1}^p\min
		\left\{\left(\sigma _{ii}\sigma _{jj}\right)^{1/2},\frac{\left\vert \sigma _{ij}\right\vert
		}{\sqrt{(\ln p)/n}}\right\}\leq c_{n,p}\right\}.
	\end{equation}
	The parameter space $\mathcal{H}(c_{n,p})$ contains a large collection of sparse covariance matrices and does not impose any constraint on the variances $\sigma _{ii}$, $i=1, ..., p$. The collection $\mathcal{H}(c_{n,p})$ contains some other commonly used classes of sparse covariance matrices in the literature, including an $\ell_{q}$ ball assumption $\max_{i}\sum_{j=1}^{p}\left\vert \sigma _{ij}\right\vert ^{q}\leq s_{n,p}$
	in Bickel and Levina \cite{Bickel2}, and a weak $\ell_{q} $ ball assumption $\max_{1\leq j\leq p}\left\{\left\vert \sigma _{j\left[ k\right] }\right\vert ^{q}\right\} \leq s_{n,p}/k $ for each integer $k$ in Cai and Zhou \cite{Cai_Zhou_covariance} where $\left\vert \sigma _{j \left[ k\right] }\right\vert $ is the $k$th largest entry in magnitude of the $j$th row $\left( \sigma _{ij}\right) _{1\leq i\leq p}$. See Cai et al. \cite{Cai_Ren_Zhou_survey} for more discussions. 

	We have the following result on the performance of $\hat\SSigma^{\rm at}$ over the parameter space $\mathcal{H}(c_{n, p})$.
	
	\begin{Theorem}\label{th:sparse_cov}
		Suppose that $\delta\geq 2$,  $\ln p = o((n_{\min}^{\ast})^{1/3})$  and  Assumptions \ref{as:MCR} and  \ref{as:sub-Gaussian} hold. Then, conditioning on $\S$,  there exists a constant  $C$ not depending on $p$, $n_{\min}^\ast$ or $n$ such that for any $\SSigma\in \mathcal{H}(c_{n, p})$, 
		\begin{equation}\label{ineq:sparse_cov_hp}
			\Pr\left(\left\|\hat{\SSigma}^{\rm at} - \SSigma\right\|\leq Cc_{n, p} \sqrt{\frac{\ln p}{n_{\min}^\ast}}\right) \geq 1 - O\left\{(\ln p)^{-1/2}p^{-\delta+2}\right\}.
		\end{equation}
		Moreover, if we further assume that $p\geq (n_{\min}^\ast)^\xi$ and $\delta \geq 4 + 1/\xi$, we in addition have
		\begin{equation}\label{ineq:sparse_cov_E}
			\Ex\left(\|\hat \SSigma^{\rm at} - \SSigma\|^2\right) \leq C c_{n, p}^2\frac{\ln p}{ n_{\min}^\ast}.
		\end{equation}
	\end{Theorem}	
	Moreover, the lower bound result below shows that the rate in \eqref{ineq:sparse_cov_E} is optimal. 
	\begin{Proposition}\label{pr:sparse_lower} For any $n_0\geq 1$ and $c_{n, p}>0$ such that $c_{n, p} \leq Mn_0^{1/2}(\ln p)^{-3/2}$ for some constant $M>0$, conditioning on $\S$ we have
				$$ \inf_{\hat{\SSigma}}\sup_{\substack{\SSigma \in \mathcal{H}(c_{n, p})\\ \S: n_{\min}^\ast \geq n_0}} \Ex \left(\|\hat{\SSigma} - \SSigma\|^2\right) \geq C c_{n, p}^2\frac{\ln p}{ n_0}. $$ 
	\end{Proposition}


	\begin{Remark}[$\ell_q$ norm loss]{\rm
			We focus in this paper on estimation under the spectral norm loss. The results given in Theorem \ref{th:sparse_cov} can be easily generalized to the general matrix $\ell_q$ norm for $1\le q \le \infty$. The results given in Equations \eqref{ineq:sparse_cov_hp} and \eqref{ineq:sparse_cov_E} remain valid when the spectral norm is replaced by the matrix $\ell_q$ norm for $1\le q \le \infty$.
		}
	\end{Remark}
	
	\begin{Remark}[Positive definiteness]{\rm
			Under mild conditions on $\SSigma$, the estimator $\hat \SSigma^{\rm at}$ is positive definite with high probability. However, $\hat \SSigma^{\rm at}$ is not guaranteed to be positive definite  for a given data set. 
			Whenever $\hat \SSigma^{\rm at}$  is not positive semi-definite, a simple extra step can make the final estimator 
			$\hat \SSigma^{\rm at}_{+}$ positive definite and also rate-optimal.
			
			Write the eigen-decomposition of  $\hat \SSigma^{\rm at}$   as $\hat \SSigma^{\rm at}=\sum_{i=1}^{p}\hat{\lambda}_{i}\hat{\v}_{i}\hat{\v}_{i}^{\top}$,
			where $\hat{\lambda}_{1} \geq \cdots \geq \hat{\lambda}_{p}$ are the eigenvalues and $\hat{\v}_{i}$ are the corresponding eigenvectors. Define the final estimator
			\begin{equation*}
				\hat \SSigma^{\rm at}_{+}=\hat \SSigma^{\rm at} + \left(|\hat{\lambda}_{p}| + {\ln p \over n_{\min}^\ast}\right) I\{\hat{\lambda}_{p}<0\}\cdot \I_{p\times p},
			\end{equation*}
			where $\I_{p\times p}$ is the $p\times p$ identity matrix. Then $\hat \SSigma^{\rm at}_{+}$ is a positive definite matrix with the same structure as that of $\hat \SSigma^{\rm at}$. It is easy to show that $\hat \SSigma^{\rm at}_{+}$ and  $\hat \SSigma^{\rm at}$ attains the same rate of convergence over $\mathcal{H}(c_{n, p})$.  See Cai, Ren and Zhou \cite{Cai_Ren_Zhou_survey} for further discussions.
		}
	\end{Remark}
	
	\begin{Remark}[Exclusion of  the Rademacher Distribution]\label{Rademacher.rmk} \rm
		To guarantee that $\hat{\theta}_{ij}^\ast$ is a good estimate of $\theta_{ij}$, one important condition needed in the theoretical analysis is that $\theta_{ij}/\sqrt{\sigma_{ii}\sigma_{jj}}$ is bounded from below by a positive constant. However when the components of $\Z_k$ in \eqref{SubGaussian-Z} are i.i.d. Rademacher, it is possible that $\theta_{ij}/\sqrt{\sigma_{ii}\sigma_{jj}} = 0$. For example, If $Z_1$ and $Z_2$ are i.i.d. Rademacher and $X_i = Z_1 + Z_2$ and $X_j = Z_1 - Z_2$, then $\var(X_iX_j) = \var(Z_1^2-Z_2^2) = 0$, and this implies $\theta_{ij}/\sqrt{\sigma_{ii}\sigma_{jj}} = 0$. 
	\end{Remark}

	A key technical tool in the analysis of the adaptive thresholding estimator is a large deviation result for the self-normalized entries of the generalized sample covariance matrix. The following lemma, proved in Section \ref{proof.sec},  plays  a critical role in the proof of Theorem \ref{th:sparse_cov} and can be useful for other high-dimensional inference problems with missing data.
	\begin{Lemma}\label{lm:covariance}
		Suppose $\ln p = o((n_{\min}^\ast)^{1/3})$ and  Assumptions \ref{as:MCR} and  \ref{as:sub-Gaussian} hold. For any constants $\delta\geq 2$, $\varepsilon>0$, $M>0$, conditioning on $\S$, we have
		\begin{equation}\label{ineq:sigma}
			\Pr\left({|\hat{\sigma}_{ij}^\ast - \sigma_{ij}|\over (\hat{\theta}_{ij}^\ast)^{1/2}} \geq \delta\sqrt{\ln p\over n^\ast_{ij}}, \forall 1\leq i,j \leq p \right) = O\left\{(\ln p)^{-1/2}p^{-\delta+2}\right\},
		\end{equation}
		\begin{equation}\label{ineq:theta}
			\Pr\left(\max_{ij}{|\hat{\theta}_{ij}^\ast - \theta_{ij}|\over \sigma_{ii}\sigma_{jj}} \geq \varepsilon \right) = O(p^{-M}).
		\end{equation}
	\end{Lemma}
	
	In addition to optimal estimation of a sparse covariance matrix $\SSigma$ under the spectral norm loss, it is also of significant interest to recover the support of $\SSigma$, i.e., the locations of the nonzero entries of $\SSigma$. The problem has been studied in the case of complete data in, e.g., Cai and Liu \cite{Cai_Liu} and Rothman et al. \cite{Rothman_sparse_cov}. With incomplete data, the support can be similarly recovered through adaptive thresholding. Specifically,  define the support of $\SSigma = (\sigma_{ij})_{1\leq i, j\leq p}$ by $\supp(\SSigma) =  \{(i, j): \sigma_{ij} \neq 0\}$.  Under the condition that the non-zero entries of $\SSigma$ are sufficiently bounded away from zero, the adaptive thresholding estimator $\hat{\SSigma}^{\rm at}$ recovers the support $\supp(\SSigma)$ consistently. It is noteworthy that in the support recovery analysis, the sparsity assumption is not directly needed.
	\begin{Theorem}[Support Recovery]\label{th:support}
		Suppose $\ln p = o((n_{\min}^\ast)^{1/3})$  and  Assumptions \ref{as:MCR} and  \ref{as:sub-Gaussian} hold. Let $\gamma$ be any positive constant. Suppose $\SSigma$ satisfies
		\begin{equation}
			|\sigma_{ij}| > (4 + \gamma) \sqrt{\frac{\theta_{ij}\ln p}{n_{ij}^\ast}}, \quad \text{ for all } (i,j) \in \supp(\SSigma).
		\end{equation}
		Let $\hat{\SSigma}^{\rm at}$ be the adaptive thresholding estimator with $\delta = 2$, then, conditioning on $\S$, we have
		\begin{equation}
			\Pr\left\{\supp (\hat\SSigma^{\rm at}) = \supp (\SSigma) \right\} \to 1 \quad \text{as}\quad n, \; p\to \infty.
		\end{equation}
	\end{Theorem}

	\section{Numerical Results}
	\label{simulation.sec}
	
	We investigate in this section the numerical performance of the proposed estimators  through simulations. 
	The proposed adaptive thresholding procedure is also illustrated with an estimation of the covariance matrix based on data from four ovarian cancer studies.
	
	The estimators $\hat{\SSigma}^{\rm bt}$ and $\hat{\SSigma}^{\rm at}$   introduced in the previous sections all require specification of the tuning parameters ($k$ or $\delta$).  Cross-validation is a simple and practical  data-driven  method for the selection of these tuning parameters. Numerical results indicate that the proposed estimators with the tuning parameter selected by cross-validation perform well empirically. We begin by introducing the following $K$-fold cross-validation method for the empirical selection of the tuning parameters.
	
	\subsection{Cross-validation}
	\label{CV.sec}
	
	
	For a pre-specified positive integer $N$,  we construct a grid $T$ of non-negative numbers.
	For bandable covariance matrix estimation, we set
	$T = \left\{1, \lceil p^{1/N} \rceil , \ldots, \lceil p^{N/N}\rceil \right\}$, and for sparse covariance matrix estimation, we let  $T = \left\{0, 1/N, \ldots, 4N/N\right\}$.
	
	Given $n$ samples $\X^\ast \in \mathbb{R}^{p\times n}$ with missing values, for a given positive integer $K$, we randomly divide them into two groups of size $n_1 \approx n(K-1)/K$, $n_2\approx n/K$  for $H$ times. For $h=1,\ldots, H$, we denote by $J_1^h$ and $J_2^h \subseteq\{1, \ldots, n\}$ the index sets of the two groups for the $h$-th split. The proposed estimator,  $\hat{\SSigma}^{\rm bt}$ for bandable covariance matrices, or $\hat{\SSigma}^{\rm at}$ for sparse covariance matrices, is then applied to the first group of data $\X_{J_1^h}^\ast$ with each value of the tuning parameter $t \in T$ and denote the result by $\hat\SSigma^{\rm bt}_h(t)$ or $\hat\SSigma^{\rm at}_h(t)$ respectively. Denote the generalized sample covariance matrix of the second group of data $\X_{J_2^h}^\ast$ by $\hat{\SSigma}_h^\ast$ and set
	\beq
	\hat{R}(t) =\frac{1}{H} \sum_{h=1}^H \| \hat{\SSigma}_h(t) - \hat{\SSigma}^\ast_{h} \|_F^2, 
	\label{CV}
	\eeq
	where $\hat{\SSigma}_h(t)$ is either $\hat{\SSigma}^{\rm bt}(t)$ for bandable covariance matrices, or $\hat{\SSigma}^{\rm at}(t)$ for sparse covariance matrices.
	
	The final tuning parameter is chosen to be
	$$t_\ast = \argmin_{T} \hat R(t) $$
	and the final estimator $\hat{\SSigma}^{\rm bt}$ (or $\hat{\SSigma}^{\rm at}$) is calculated using this choice of the tuning parameter  $t_\ast$.
	In the following numerical studies, we will use 5-fold cross-validation (i.e., $K=5$) to select the tuning parameters.
	
	\begin{Remark}\rm
		The Frobenius norm used in \eqref{CV} can be replaced by other losses such as the spectral norm. Our simulation results indicate that using the Frobenius norm in \eqref{CV} works well, even when the true loss is the spectral norm loss.
	\end{Remark}
	
	\subsection{Simulation Studies}
	
	In the simulation studies, we consider the following two settings for the missingness. The first is MUCR where each entry $X_{ik}$ is observed with probability $0< \rho \le 1$, and the second is
	missing not uniformly but completely at random (MCR)  where the complete data matrix $\X$ is divided into four equal-size blocks,
	$$\X = \begin{bmatrix}
	\X_{(11)} & \X_{(12)}\\
	\X_{(21)} & \X_{(22)}\\
	\end{bmatrix}, \quad \X_{(11)}, \X_{(12)}, \X_{(21)}, \X_{(22)}\in \mathbb{R}^{\frac{p}{2}\times \frac{n}{2}},$$
	and each entry of  $\X_{(11)}$ and $\X_{(22)}$ is observed with probability $\rho^{(1)}$ and each entry of $\X_{(12)}$ and $\X_{(21)}$ is observed with probability $\rho^{(2)}$, for some $0< \rho^{(1)}, \rho^{(2)} \le 1$.

	As mentioned in the introduction, high-dimensional inference for missing data has been studied in the case of MUCR and we would like to compare our estimators with the corresponding estimators based on a different sample covariance matrix designed for the MUCR case.  Under the assumption that $\Ex\X = 0$ and each entry of $\X$ is observed independently with probability $\rho$, Wainwright \cite{Loh} and Lounici \cite{Lounici2} introduced the following substitute of the usual sample covariance matrix
	\begin{equation}\label{eq:bullet_1}
		\hat{\SSigma}^{\bullet} = (\sigma_{ij}^\bullet)_{1\leq i, j\leq p}\quad \mbox{with}\quad  \hat{\sigma}_{ij}^\bullet = \left\{ \begin{array}{ll}
			\frac{1}{n(1-\rho)^2}\sum_{k=1}^n X_{ik}^\ast X_{jk}^\ast, & i\neq j \vspace*{1em}\\
			\frac{1}{n(1-\rho)}\sum_{k=1}^n X_{ik}^\ast X_{jk}^\ast, & i=j
		\end{array} \right.
	\end{equation}
	where the missing entries of $\X^\ast$ are replaced by 0's. 
	It is easy to show that $\hat{\SSigma}^\bullet$ is a consistent estimator of $\SSigma$ under MUCR and could be used similarly as the sample covariance matrix in the complete data setting. 
	
	For more general settings where $\Ex\X\neq 0$ and the coordinates $X_1,  X_2, ..., X_p$ are observed with different probabilities $\rho_1, \ldots, \rho_p$, $\hat{\SSigma}^\bullet$ can be generalized as 
	\begin{equation}\label{eq:bullet_2}
		\hat{\SSigma}^\bullet = (\hat\sigma^\bullet_{ij})_{1\leq i,j\leq p}\quad \mbox{with}\quad  \hat{\sigma}^\bullet_{ij} = \left\{ \begin{array}{ll}
			\frac{1}{n(1-\hat\rho_i)(1-\hat\rho_j)}\sum_{k=1}^n X_{ik, c}^\ast X_{jk,c }^\ast, & i\neq j \vspace*{1em}\\
			\frac{1}{n(1-\hat\rho_i)}\sum_{k=1}^n X_{ik,c}^\ast X_{jk,c}^\ast, & i=j
		\end{array} \right.  
	\end{equation}
	where for $i=1,\ldots, p$ and $k = 1,\ldots, n$,  $\hat \rho_i = \frac{1}{n}\sum_{k=1}^n S_{ik}$  and
	$X^\ast_{ik, c} = X^\ast_{ik} - \bar X_i^\ast$ . 
	
	Based on $\hat{\SSigma}^\bullet$, we can analogously define the corresponding blockwise tridiagonal estimator $\hat{\SSigma}^{\rm bt\bullet}$ for bandable covariance matrices, and adaptive thresholding estimator $\hat{\SSigma}^{\rm  at\bullet}$ for sparse covariance matrices.
	
	We first consider estimation of bandable covariance matrices and compare the proposed blockwise tridiagonal estimator $\hat{\SSigma}^{\rm bt}$ with  the corresponding estimator $\hat{\SSigma}^{\rm bt \bullet}$. For both methods, the tuning parameter $k$ is selected by 5-fold cross-validation with $N$ varying from 20 to 50. The following bandable covariance matrices are considered:
	\begin{enumerate}
		\item (Linear decaying bandable model) $\SSigma = (\sigma_{ij})_{1\leq i, j\leq p}$ with $\sigma_{ij} = \max\{0, 1 - |i-j|/5\}$.
		
		\item (Squared decaying bandable model) $\SSigma = (\sigma_{ij})_{1\leq i, j\leq p}$ with $\sigma_{ij} = (|i-j| + 1)^{-2}$.
	\end{enumerate}
	For missingness,  both MUCR and MCR are considered and \eqref{eq:bullet_1} and \eqref{eq:bullet_2} are used to calculate $\hat{\SSigma}^\bullet$ respectively. 
	The proposed procedure $\hat{\SSigma}^{\rm bt}$ is compared with  the estimator $\hat{\SSigma}^{\rm bt\bullet}$, which is based on $\hat{\SSigma}^{\bullet}$. The results for the spectral norm, $\ell_1$ norm and Frobenius norm losses are reported in Table \ref{tb:compare_bandable}. It is easy to see from Table \ref{tb:compare_bandable} that the proposed estimator $\hat{\SSigma}^{\rm bt}$ generally outperforms $\hat{\SSigma}^{\rm bt\bullet}$, especially in the fast decaying setting.
	
	\begin{table} 
		\begin{tabular}{ccccccc}
			\hline
			& \multicolumn{2}{c}{Spectral norm}  & \multicolumn{2}{c}{$\ell_1$ norm} & \multicolumn{2}{c}{Frobenius norm}\\
			\cmidrule(lr){2-3}\cmidrule(lr){4-5}\cmidrule{6-7}
			$(p, n)$ & $\hat{\SSigma}^{\rm bt}$ & $\hat{\SSigma}^{\rm bt\bullet}$ & $\hat{\SSigma}^{\rm bt}$ & $\hat{\SSigma}^{\rm bt\bullet}$ & $\hat{\SSigma}^{\rm bt}$ & $\hat{\SSigma}^{\rm bt\bullet}$ \\\hline 
			\multicolumn{7}{c}{Linear Decay Bandable Model, MUCR $\rho = .5$}\\
			$(50, 50)$& 2.78(0.17) & 2.88(0.18) & 4.37(0.57) & 4.57(0.76) & 7.73(0.85) & 7.85(0.80)\\
			$(50, 200)$& 1.44(0.06) & 1.56(0.07) & 2.52(0.17) & 2.71(0.19) & 3.91(0.18) & 4.16(0.16)\\
			$(200, 100)$& 2.25(0.13) & 2.44(0.16) & 3.83(0.32) & 4.22(0.46) & 10.27(0.29) & 10.89(0.29)\\
			$(200, 200)$& 1.67(0.07) & 1.82(0.08) & 2.81(0.19) & 3.08(0.22) & 7.19(0.19) & 7.68(0.14)\\
			$(500, 200)$& 2.00(0.07) & 2.18(0.10) & 3.45(0.16) & 3.74(0.27) & 12.10(0.36) & 12.87(0.42)\\
			\multicolumn{7}{c}{Squared Decay Bandable Model, MUCR $\rho = .5$}\\
			$(50, 50)$& 1.34(0.08) & 1.40(0.11) & 2.28(0.16) & 2.37(0.21) & 3.78(0.19) & 3.91(0.18)\\
			$(50, 200)$& 0.82(0.01) & 0.84(0.01) & 1.47(0.03) & 1.49(0.02) & 2.24(0.02) & 2.30(0.02)\\
			$(200, 100)$& 1.13(0.01) & 1.17(0.02) & 2.12(0.05) & 2.18(0.07) & 5.74(0.04) & 5.91(0.05)\\
			$(200, 200)$& 0.92(0.00) & 0.94(0.00) & 1.66(0.02) & 1.72(0.03) & 4.49(0.02) & 4.61(0.01)\\
			$(500, 200)$& 0.97(0.00) & 0.98(0.00) & 1.80(0.02) & 1.86(0.02) & 7.15(0.01) & 7.35(0.01)\\
			\multicolumn{7}{c}{Linear Decay Bandable Model, MCR $\rho^{(1)} = .8, \rho^{(2)} = .2$}\\
			$(50, 50)$& 2.76(0.26) & 3.46(1.43) & 4.24(0.73) & 5.87(2.91) & 7.03(1.25) & 8.47(1.29)\\
			$(50, 200)$& 1.51(0.11) & 2.64(0.40) & 2.52(0.30) & 4.29(0.99) & 3.62(0.30) & 5.77(0.45)\\
			$(200, 100)$& 2.32(0.22) & 3.93(0.67) & 3.73(0.47) & 6.21(1.11) & 9.04(0.48) & 13.47(0.84)\\
			$(200, 200)$& 1.67(0.10) & 3.23(0.27) & 2.71(0.26) & 4.91(0.49) & 6.32(0.11) & 11.32(0.49)\\
			$(500, 200)$& 1.98(0.09) & 3.78(0.20) & 3.19(0.20) & 5.70(0.42) & 10.39(0.12) & 18.48(0.49)\\
			\multicolumn{7}{c}{Squared Decay Bandable Model, MCR $\rho^{(1)} = .8, \rho^{(2)} = .2$}\\
			$(50, 50)$& 1.26(0.08) & 1.49(0.13) & 2.21(0.23) & 2.60(0.28) & 3.48(0.14) & 4.18(0.23)\\
			$(50, 200)$& 0.82(0.01) & 0.88(0.04) & 1.47(0.05) & 1.77(0.11) & 2.18(0.04) & 2.68(0.11)\\
			$(200, 100)$& 1.06(0.01) & 1.30(0.04) & 1.96(0.04) & 2.44(0.07) & 5.32(0.02) & 6.51(0.06)\\
			$(200, 200)$& 0.90(0.00) & 0.96(0.03) & 1.60(0.02) & 1.99(0.06) & 4.27(0.02) & 5.26(0.15)\\
			$(500, 200)$& 0.93(0.00) & 1.03(0.01) & 1.69(0.01) & 2.11(0.03) & 6.73(0.01) & 8.25(0.04)\\
			\hline
		\end{tabular}
		\caption{Comparsion between $\hat{\SSigma}^{\rm bt}$ and $\hat{\SSigma}^{\rm bt\bullet}$ in different settings of bandable covariance matrix estimation.}
		\label{tb:compare_bandable}
	\end{table}

	Now we consider estimation of sparse covariance matrices with missing values under the following two models.
\begin{enumerate}
	\item (Permutation Bandable Model) $\SSigma = (\sigma_{ij})_{1\leq i, j\leq p}$, where $\sigma_{ij} = \max(0, 1 - 0.2\cdot |s(i)-s(j)|)$ and $s(i), i=1,\dots, p$ is a random permutation of $\{1, \ldots, p \}$.
	\item (Randomly Sparse Model) $\SSigma = \I_p + (\D + \D\trans)/(\|\D + \D\trans\|+0.01) $, where $\D$ is randomly generated as
	$$\D = (d_{ij})_{1\leq i, j\leq p}, \quad d_{ij} = \left\{\begin{array}{ll}
	1 & \text{w.p. } 0.1\\
	0 & \text{w.p. } 0.8 \\
	-1 & \text{w.p. } 0.1\\
	\end{array}\right. \quad \text{for $i\neq j$};\quad d_{ii} = 0. $$
\end{enumerate}
Similar to the sparse covariance matrix estimation, for missingness, we consider both MUCR and MCR. The results for the spectral norm, matrix $\ell_1$ norm and Frobenius norm losses are summarized in Table \ref{tb:compare}.
It can be seen from Table \ref{tb:compare} that, even under the MUCR setting, the proposed estimator $\hat{\SSigma}^{\rm at}$ based on the generalized sample covariance matrix is uniformly better than the one based on $\hat{\SSigma}^\bullet$. 
In the more general MCR setting, the difference in the performance between the two estimators is even more significant.

%

\begin{table} 
	\begin{center}
		\begin{tabular}{ccccccc}
			\hline
			& \multicolumn{2}{c}{Spectral norm}  & \multicolumn{2}{c}{$\ell_1$ norm} & \multicolumn{2}{c}{Frobenius norm}\\
			\cmidrule(lr){2-3}\cmidrule(lr){4-5}\cmidrule{6-7}
			$(p, n)$ & $\hat{\SSigma}^{\rm at}$ & $\hat{\SSigma}^{\rm at\bullet}$ & $\hat{\SSigma}^{\rm at}$ & $\hat{\SSigma}^{\rm at\bullet}$ & $\hat{\SSigma}^{\rm at}$ & $\hat{\SSigma}^{\rm at \bullet}$ \\\hline 
			\multicolumn{7}{c}{Permutation Bandable Model, MUCR $\rho = .5$}\\
			$(50, 50)$& 4.26(0.24) & 4.45(0.41) & 5.58(0.58) & 6.19(7.54) & 11.34(0.79) & 11.73(1.08)\\
			$(50, 200)$& 1.70(0.05) & 1.74(0.06) & 3.31(0.32) & 3.42(0.38) & 4.93(0.09) & 5.07(0.16)\\
			$(200, 100)$& 3.48(0.07) & 3.66(0.58) & 5.80(0.39) & 6.23(14.89) & 18.34(0.81) & 19.37(5.50)\\
			$(200, 200)$ & 2.12(0.04) & 2.20(0.03) & 4.17(0.29) & 4.44(0.32) & 11.46(0.14) & 11.94(0.13) \\
			$(500, 200)$& 2.28(0.03) & 3.51(0.17) & 4.17(0.15) & 6.55(0.72) & 16.85(0.10) & 21.96(0.49)\\
			\multicolumn{7}{c}{Randomly Sparse Model, MUCR $\rho = .5$}\\
			$(50, 50)$& 1.76(0.07) & 1.96(0.62) & 3.69(0.24) & 4.20(5.89) & 5.75(0.51) & 6.27(2.95)\\
			$(50, 200)$& 1.05(0.00) & 1.06(0.00) & 2.73(0.04) & 2.74(0.05) & 3.75(0.03) & 3.77(0.04)\\
			$(200, 100)$& 1.40(0.01) & 1.45(0.01) & 4.88(0.08) & 4.94(0.09) & 8.34(0.07) & 8.50(0.07)\\
			$(200, 200)$& 1.07(0.00) & 1.09(0.01) & 4.44(0.03) & 4.46(0.03) & 7.42(0.02) & 7.43(0.02)\\
			$(500, 200)$& 1.14(0.01) & 1.31(0.01) & 6.39(0.04) & 6.65(0.08) & 11.73(0.01) & 12.23(0.05)\\
			\multicolumn{7}{c}{Permutation Bandable Model, MCR $\rho^{(1)} = .8, \rho^{(2)} = .2$}\\
			$(50, 50)$& 4.23(0.38) & 4.71(1.17) & 6.67(2.30) & 7.46(8.92) & 11.22(1.34) & 11.71(2.01)\\
			$(50, 200)$& 1.64(0.05) & 2.79(0.39) & 2.94(0.21) & 4.52(0.95) & 4.41(0.13) & 6.29(0.46)\\
			$(200, 100)$& 3.17(0.06) & 4.16(0.57) & 5.73(0.66) & 8.11(1.87) & 15.93(0.53) & 18.03(0.77)\\
			$(200, 200)$& 2.00(0.03) & 3.22(0.18) & 3.65(0.16) & 5.70(0.60) & 9.83(0.11) & 13.29(0.55)\\
			$(500, 200)$& 2.22(0.03) & 3.45(0.17) & 4.09(0.17) & 6.44(0.96) & 16.80(0.14) & 21.93(0.45)\\
			\multicolumn{7}{c}{Randomly Sparse Model, MCR $\rho^{(1)} = .8, \rho^{(2)} = .2$}\\
			$(50, 50)$& 2.15(0.46) & 2.19(0.49) & 4.21(0.94) & 4.47(4.65) & 6.36(0.96) & 7.25(1.57)\\
			$(50, 200)$& 1.09(0.02) & 1.16(0.04) & 2.82(0.19) & 2.99(0.32) & 3.83(0.10) & 4.00(0.20)\\
			$(200, 100)$& 1.46(0.02) & 1.82(0.03) & 4.96(0.12) & 5.61(0.21) & 8.45(0.07) & 10.10(0.14)\\
			$(200, 200)$& 1.08(0.00) & 1.20(0.01) & 4.46(0.04) & 4.57(0.05) & 7.43(0.02) & 7.66(0.04)\\
			$(500, 200)$& 1.12(0.01) & 1.33(0.01) & 6.35(0.04) & 6.60(0.07) & 11.71(0.02) & 12.20(0.06)\\\hline
		\end{tabular}		
		\caption{Comparsion between $\hat{\SSigma}^{\rm at}$ and $\hat{\SSigma}^{\rm at\bullet}$ in different settings of sparse covariance matrix estimation.}
		\label{tb:compare}
	\end{center}
\end{table}

	\subsection{Comparison with Complete Samples}
	
	For covariance matrix estimation with missing data, an interesting question is: what is the ``effective sample size"? That is, for samples with missing values, we would like to know the equivalent size of complete samples such that the accuracy for covariance matrix estimation is approximately the same. 
	We now compare the performance of the proposed estimator based on the incomplete data with the corresponding estimator based on  the complete data for various sample sizes. We fix the dimension $p=100$. For the incomplete data, we consider $n = 1000$ and MUCR with $\rho = .5$. The covariance matrix $\SSigma$ is chosen as 
	\begin{itemize}
		\item \text{Linear Decaying Bandable Model}  \text{(in Bandable Covariance Matrix Estimation);}
		\item \text{Permutation Bandable Model}  \text{(in Sparse Covariance Matrix Estimation);}
	\end{itemize}
	Correspondingly, we consider the similar settings for the complete data with the same  $\SSigma$ and $p$, but different sample size $n_c$, where $n_c$ can be one of the following three values,
	\begin{enumerate}
		\item $\overline{n_{\rm pair}^\ast} =  \sum_{i,j=1}^n n_{ij}^\ast/p^2$:  the average number of pairs of $(x_i, x_j)$'s that can be observed within the same sample;
		\item $\overline{n_{\rm s}^\ast} = \sum_{i=1}^n n_{i}^\ast/p$: the average number of single $x_i$'s can be observed;
		\item $n$: the same number of samples with the missing values.
	\end{enumerate}
	The results for all the settings are summarized in Table \ref{tb:compare_complete}. It can be seen that the equivalent sample size depends on the loss function and in general it is between $\overline{n_{\rm pair}^\ast} $ and $\overline{n_{\rm s}^\ast}$. 
	Overall, the average risk under the missing data setting is most comparable to that under the complete data setting for the sample size of $n_c=\overline{n_{\rm pair}^\ast}$, the average number of observed pairs.
	
	\begin{table} 
		\begin{center}
		\begin{tabular}{rcccc}
			\hline
			Setting & sample size & Spectral norm & $\ell_1$ norm & Frobenius norm\\\hline
			\multicolumn{5}{c}{Bandable Covariance Matrix Estimation}\\
			Missing Data & $n = 1000$ & 0.72(0.01)  & 1.25(0.03) & 2.40(0.01) \\
			Complete Data & $n_c=\overline {n_{\rm pair}^\ast}$ & 0.97(0.03)  & 1.49(0.05) & 2.48(0.04)\\
			Complete Data & $n_c=\overline{n_{\rm s}^\ast}$ & 0.65(0.01) & 1.01(0.03) & 1.69(0.03) \\
			Complete Data & $n_c=n$ & 0.48(0.01) & 0.73(0.01) & 1.22(0.01)\\\hline 
			\multicolumn{5}{c}{Sparse Covariance Matrix Estimation}\\
			Missing Data & $n = 1000$ & 0.75(0.01) & 1.37(0.04) & 2.90(0.02) \\
			Complete Data & $n_c=\overline {n_{\rm pair}^\ast}$ & 0.83(0.02) & 1.31(0.05)  & 2.94(0.04)\\
			Complete Data & $n_c=\overline{n_{\rm s}^\ast}$ & 0.65(0.01)  & 1.01(0.03) & 1.86(0.04)  \\
			Complete Data & $n_c=n$ & 0.45(0.01) & 0.64(0.01) & 1.12(0.01) \\\hline
		\end{tabular}
		\caption{Comparison between incomplete samples and complete samples.}
		\label{tb:compare_complete}
		\end{center}
	\end{table}

	\subsection{Analysis of Ovarian Cancer Data}
	\label{real.data.sec}

\begin{figure}[htbp]
		\begin{center}
			\centerline{
				\includegraphics[width=6in, height=2in]{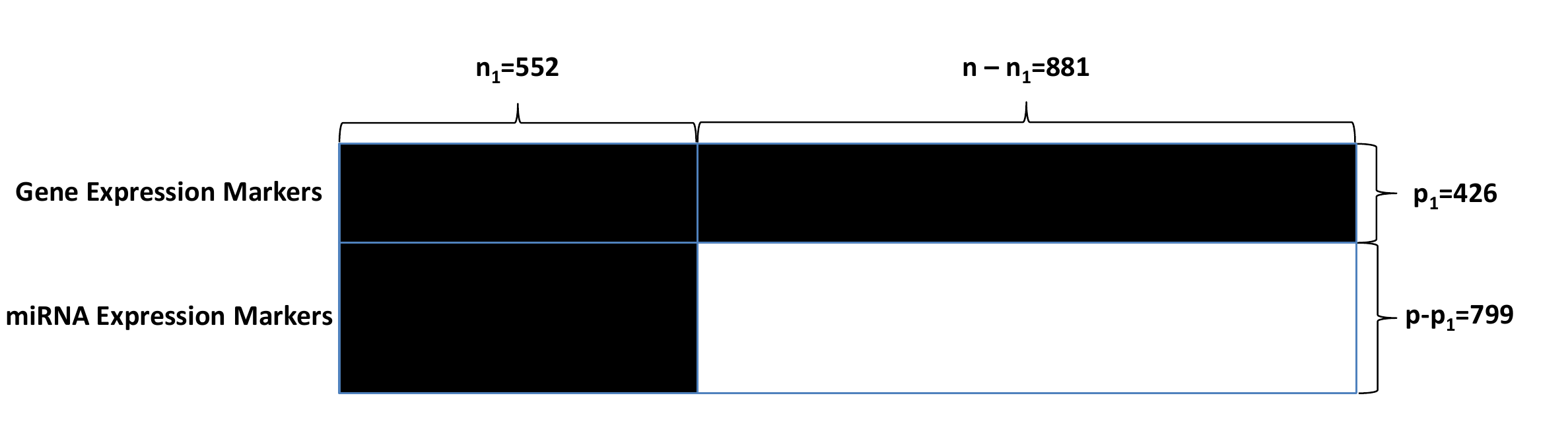}}
			\caption{Illustration of the ovarian cancer  dataset. Black block = completely observed; White block = completely missing.
			}
			\label{fig:real_data_illustration}
		\end{center}
	\end{figure}

	In this section, we illustrate the proposed adaptive thresholding procedure with an application to data from four ovarian cancer genomic studies,  Cancer Genome Atlas Research Network \cite{cancer2011integrated} (TCGA),  Bonome et al. \cite{bonome2005expression} (BONO), Dressman et al. \cite{dressman2007integrated}  (DRES) and Tothill et al. \cite{tothill2008novel}  (TOTH).
	The method introduced in Sections \ref{sparse.covariance.sec} enables us to estimate the covariance matrix by integrating data from all four studies and thus yields a more accurate estimator. 
	The data structure is illustrated in Figure \ref{fig:real_data_illustration}.  The gene expression markers (the first 426 rows) are observed in all four studies without any missingness (the top black block in Figure \ref{fig:real_data_illustration}). The miRNA expression markers are observed in 552 samples from the TCGA study (the bottom left block in Figure \ref{fig:real_data_illustration}) and completely missing in the 881 samples from  the TOTH, DRES, BONO and part of TCGA studies (the white block in Figure \ref{fig:real_data_illustration}).

\begin{figure}[htbp]
		\begin{center}
			\subfigure[Covariance matrix of the gene and miRNA expression markers. 
			The gene expression markers are marked with lines.]{\includegraphics[height = 2.7in]{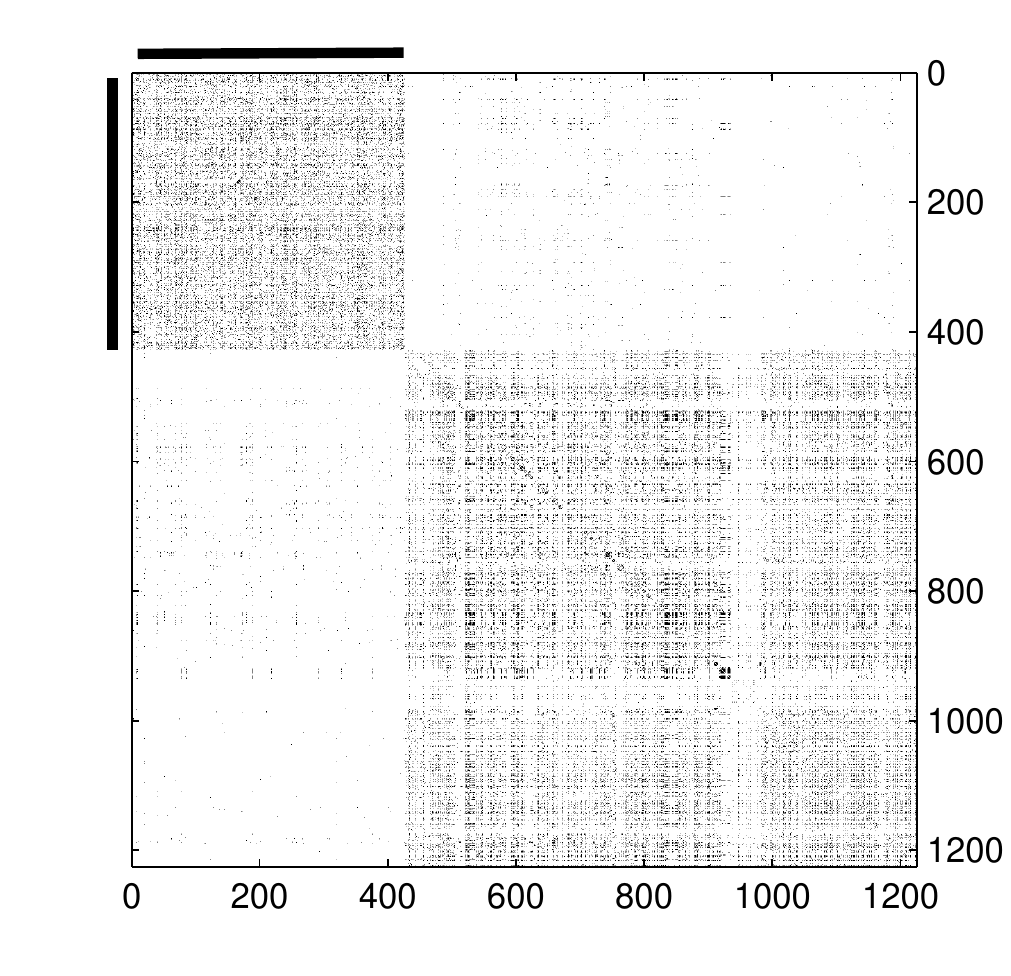}}\quad 
			\subfigure[Cross-covariances between the gene and miRNA expression markers. 1294 (.38\%) gene-miRNA pairs were detected.]{\includegraphics[height = 2.7in]{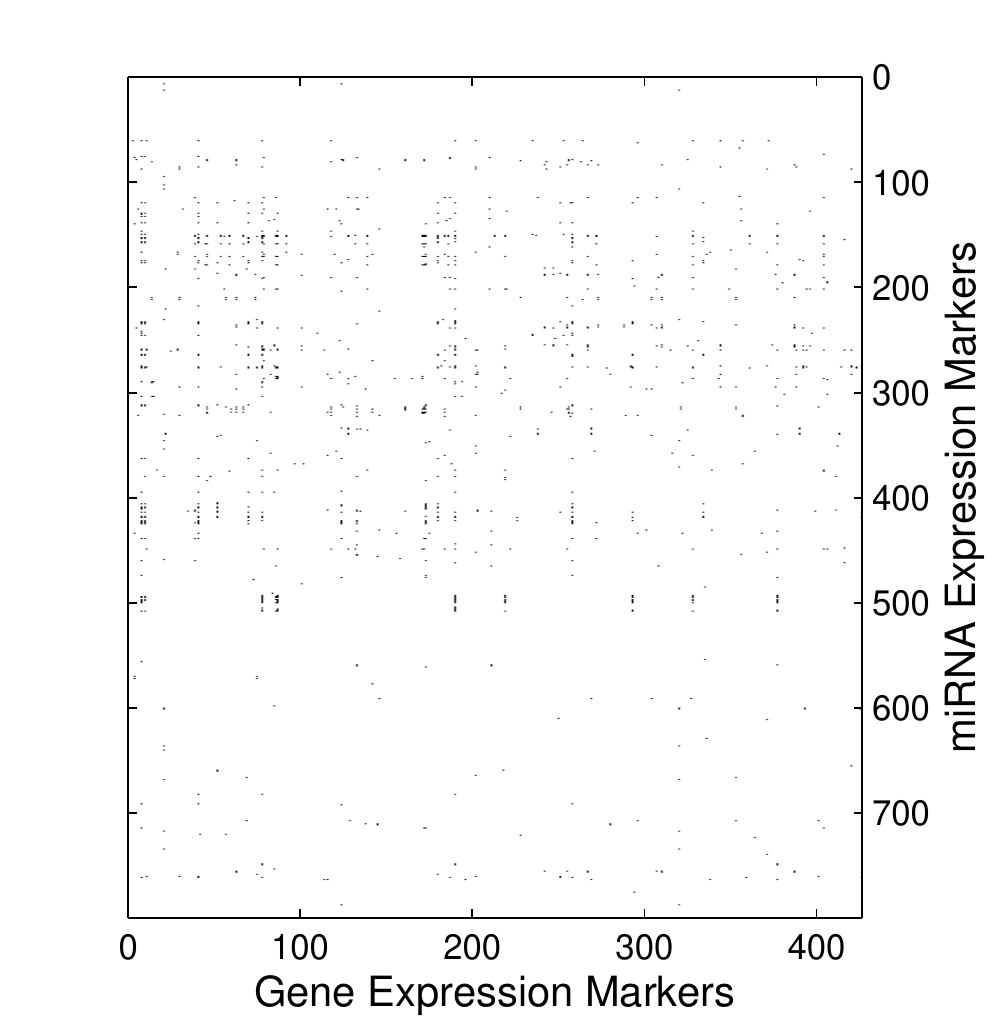}}
			\caption{Heatmaps of the covariance matrix estimate with all the observed data.
			}
			\label{fig:real_data_heatmap}
		\end{center}
	\end{figure}

	Our goal is to estimate the covariance matrix $\SSigma$ of the 1225 variables with the particular interest in the cross-covariances between the gene and miRNA expression markers. It is clear that the missingness here is not uniformly at random. On the other hand, it is reasonable to assume the missingness does not depend on the value of the data and thus missing completely at random (Assumption \ref{as:MCR}) can be assumed. 
	We apply the adaptive thresholding procedure  with $\delta = 2$ to estimate the covariance matrix and recover its support based on all the observations. The support of the estimate is shown in a heatmap in Figure \ref{fig:real_data_heatmap}. The left panel is for the whole covariance matrix and the right panel zooms into the cross-covariances between the gene and miRNA expression markers.
	
	It can be seen from Figure \ref{fig:real_data_heatmap} that the two diagonal blocks, with 12.24\% and 8.39\% nonzero off-diagonal entries respectively,  are relatively dense, indicating that the relationships among the gene expression markers and those among the miRNA expression markers, as measured by their covariances, are relatively close. 
	In contrast,  the cross-covariances between gene and miRNA expression markers are very sparse with only 0.38\% of significant gene-miRNA pairs. The gene and miRNA expression markers affect each other through different mechanisms, the cross-covariances between the gene and miRNA markers are of significant interest (see Ko et al. \cite{ACTA2}). It is worthwhile to take a closer look at the cross-covariance matrix displayed on the right panel in Figure \ref{fig:real_data_heatmap}. For each given gene, we count the number of miRNAs whose covariances with this gene are significant, and then rank all the genes by the counts. Similarly, we rank all the miRNAs. The top 5 genes and the top 5 miRNA expression markers are shown in Table \ref{tb:real_data_most_pairs}.
	
	\begin{table} 
		\begin{center}
		\begin{tabular}{ll|ll}\hline
			Gene Expression Marker & Counts & miRNA Expression Marker & Counts\\\hline
			ACTA2 & 61 & hsa-miR-142-5p & 31\\
			INHBA  & 57 &  hsa-miR-142-3p & 29\\
			COL10A1 & 53 &  hsa-miR-22 & 26 \\
			BGN  &  46 &   hsa-miR-21* & 24 \\
			NID1 & 41 &  hsa-miR-146a & 21\\\hline
		\end{tabular}
		\caption{Genes and miRNA's with most selected pairs}
		\end{center}
		\label{tb:real_data_most_pairs}
	\end{table}
	
	Many of these gene and miRNA expression markers have been studied before in the literature. For example, the miRNA expression markers hsa-miR-142-5p and hsa-miR-142-3p have been demonstrated in Andreopoulos and Anastassiou \cite{hsa-miR-142} as standing out among the miRNA markers as having higher correlations with more genes, as well as methylation sites. 
	Carraro et al. \cite{carraro2014mir} finds that  inhibition of miR-142-3p leads to ectopic expression of  the gene marker ACTA2. This indicates strong interaction between miR-142-3p and ACTA2.
	
\begin{figure}[htbp]
		\begin{center}
			\subfigure[Covariance matrix of the gene and miRNA expression markers. 
			The gene expression markers are marked with lines.]{\includegraphics[height = 2.7in]{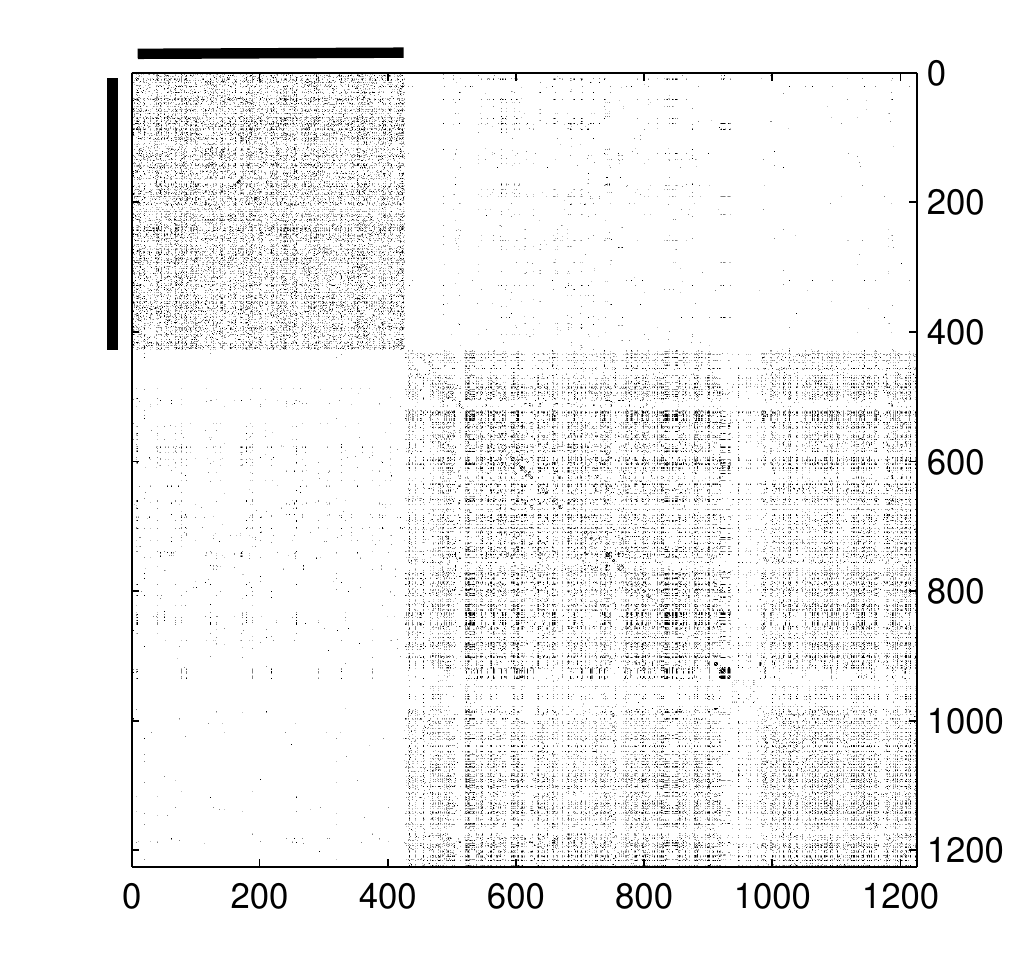}}\quad 
			\subfigure[Cross-covariances between the gene and miRNA expression markers. 1176 (.35\%)  gene-miRNA pairs were detected.]{\includegraphics[height = 2.7in]{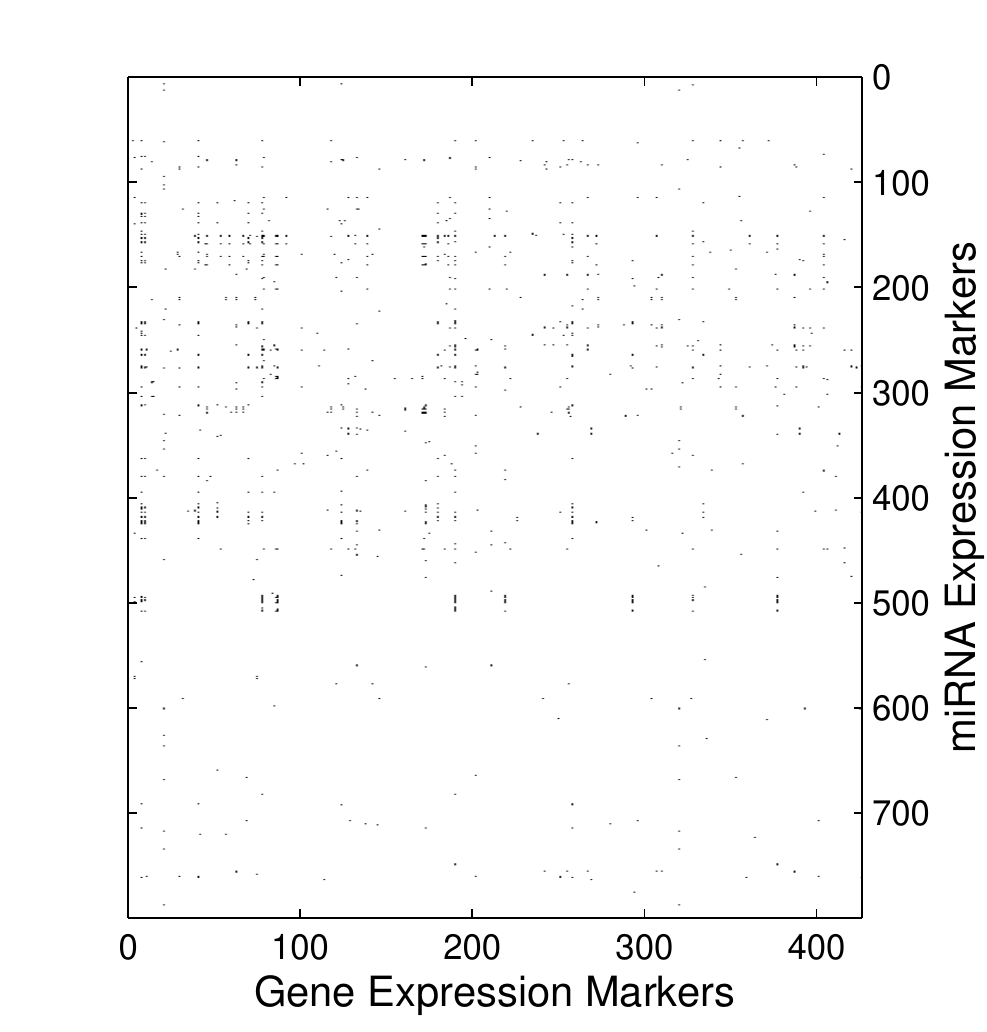}}
			\caption{Heatmaps of the covariance matrix estimate with additional missing values.
			}
			\label{fig:real_data_heatmap_missing}
		\end{center}
	\end{figure}

	To further demonstrate the robustness of our proposed procedure against missingness, we consider a setting with additional missing observations. We first randomly select half of the 552 complete samples (where both gene and miRNA expression markers are observed) and half of the 881 incomplete samples (where only gene expression markers are observed), and then independently mask each entry of the selected samples with probability $0.05$. The proposed adaptive thresholding procedure is then applied to the data with these additional missing values. The estimated covariance matrix is shown in heatmaps in Figure \ref{fig:real_data_heatmap_missing}. 
	These additional missing observations do not significantly affect the estimation accuracy.  Figure \ref{fig:real_data_heatmap_missing} is visually very similar to Figure \ref{fig:real_data_heatmap}. To quantify the similarity between the two estimates, we calculate the Matthews correlation coefficient (MCC) between them.
	The value of MCC is equal to 0.9441, which indicates that the estimate based on the data with the additional missingness is very close to the estimate based on the original samples. We also pay close attention to the cross-covariance matrix displayed on the right panel in Figure \ref{fig:real_data_heatmap_missing} and rank the gene and miRNA expression markers in the same way as before. The top 5 genes and the top 5 miRNA expression markers, listed in Table \ref{tb:real_data_most_pairs_missing}, are nearly identical to those given in Table \ref{tb:real_data_most_pairs}, which are based on the original samples.
	These results indicate that the proposed method is robust against additional missingness.
	
	\begin{table}
		\begin{center}
		\begin{tabular}{ll|ll}\hline
			Gene Expression Marker & Counts & miRNA Expression Marker & Counts\\\hline
			ACTA2 & 60 & hsa-miR-142-3p    & 31\\
			INHBA  & 56 &  hsa-miR-142-5p  & 30\\
			COL10A1 & 50 &  hsa-miR-146a   & 21 \\
			BGN  &  43 &   hsa-miR-150 & 21 \\
			NID1 & 40 &  hsa-miR-21* & 21\\\hline
		\end{tabular}
		\caption{Genes and miRNA's with most selected pairs after masking}
		\label{tb:real_data_most_pairs_missing}
		\end{center}
	\end{table}


	\section{Discussions}
	\label{discussion.sec}
	
	We considered in the present paper  estimation of bandable and sparse covariance matrices in the presence of missing observations. The pivotal quantity is the generalized sample covariance matrix defined in \eqref{eq:hat_Sigma}. 
	The technical analysis is more challenging due to the missing data. We have mainly focused on the spectral norm loss in the theoretical analysis. Performance under other losses such as the Frobenius norm can also be analyzed.
	
	To illustrate the proposed methods, we integrated four ovarian cancer studies. 
	These methods for high-dimensional covariance matrix estimation with missing data are also useful for other types of data integration. For example, linking multiple data sources such as electronic data records, medicare data, registry data and patient reported outcomes could greatly increase the power of exploratory studies such as phenome-wide association studies (Denny et al. \cite{Denny}). However, missing data inevitably arises and may hinder the potential of integrative analysis. In addition to random missingness due to unavailable information on a small fraction of patients, many variables such as the genetic measurements may only exist in one or two data sources and are hence structurally missing for other data sources. Our proposed methods could potentially provide accurate recovery of the covariance matrix in the presence of missingness. 
	
	In this paper, we allowed the proportion of missing values to be non-negligible as long as the minimum number of occurrences  of any pair of variables $n_{\min}^\ast$ is of order $n$. An interesting question is what happens when the number of observed values is large but $n_{\min}^\ast$ is  small (or even zero). We believe that  the covariance matrix $\SSigma$ can still be well estimated under certain global structural assumptions. This is out of the scope of the present paper and is an interesting problem for future research.
	
	The key ideas and techniques developed in this paper can be used for a range of other related problems in high-dimensional statistical inference with missing data. For example, the same techniques can also be applied to estimation of other structured covariance matrices such as  Toeplitz matrices, which have been studied in the literature in the case of complete data. When there are missing data, we can construct similar estimators using the  generalized sample covariance matrix. The  large deviation bounds for a sub-matrix and self-normalized entries of the generalized sample covariance matrix developed in Lemmas \ref{lm:covariance} and \ref{lm:covariance_spectral_b} would be helpful for analyzing the properties of the estimators.
	
	The techniques can also be used on two-sample problems such as estimation of differential correlation matrices and hypothesis testing on the covariance structures. The generalized sample covariance matrix can be standardized to form the generalized sample correlation matrix which can then be used to estimate the differential correlation matrix in the two-sample case. It is also of significant interest in some applications to test the covariance structures in both one- and two-sample settings based on incomplete data. 
	In the one-sample case,  it is of  interest to test the hypothesis $\{H_0: \SSigma=\I\}$  or $\{H_0: \mathbf{R}=\I\}$, where $\mathbf{R}$ is the correlation matrix. In the two-sample case, one wishes to test the equality of two covariance matrices $\{H_0:\SSigma_1 = \SSigma_2\}$.
	These are interesting problems for further exploration in the future.


	%
	
	\section{Proofs}
	\label{proof.sec}
	
	We prove Theorem \ref{th:bandable} and the key technical result Lemma \ref{lm:spectral_AB} for the bandable covariance matrix estimation in this section.

	\subsection{Proof of Lemma \ref{lm:covariance_spectral_b}} 
	
	To prove this lemma, we first introduce the following technical tool for the spectral norm of the sub-matrices.
	\begin{Lemma}\label{lm:spectral_AB}
		Suppose $\SSigma\in \mathbb{R}^{p\times p}$ is any positive semi-definite matrix, $A, B\in \{1,\ldots, p\}$, then
		\begin{equation}
			\|\SSigma_{A\times B}\| \leq (\|\SSigma_A\| \|\SSigma_B\|)^{1/2}.
		\end{equation}
	\end{Lemma}
	The proof of Lemma \ref{lm:spectral_AB} is provided later and now we move back to the proof of Lemma \ref{lm:covariance_spectral_b}. Without loss of generality, we assume that $\m = \Ex\X = 0$. We further define
	\begin{equation}
		\breve{\SSigma}^* = (\breve{\sigma}^*_{ij})_{1\leq i, j\leq p}, \quad \breve{\sigma}^*_{ij} = \frac{1}{n_{ij}^*} \sum_{k=1}^n X_{ik} X_{jk} S_{ik}S_{jk}.
	\end{equation}
	Also for convenience of presentation, we use $C, C_1, c, \ldots$ to denote uniform constants, whose exact values may vary in different senarios. The lemma is now proved in the following steps:
	\begin{enumerate}
		\item We first consider for fixed unit vectors $\a, \b\in \mathbb{R}^p$ with ${\rm supp}(\a)\subseteq A, {\rm supp}(\b)\subseteq B$, the tail bound of $\a\trans(\hat{\SSigma}^* - \SSigma)\b$. We would like to show that there exist uniform constants $C_1, c>0$ such that for all $x>0$,
		\begin{equation}\label{ineq:a(hat - Sigma)b}
			\begin{split}
				& \Pr\left\{\left|\a\trans\left(\hat{\SSigma}^\ast - \SSigma\right)\b\right|\geq x\right\}\\
				\leq & C_1\exp\left\{-cn_{\min}^\ast\min\left(\frac{x^2}{\tau^4\|\SSigma_A\|\|\SSigma_B\|}, \frac{x}{\tau^2(\|\SSigma_A\|\|\SSigma_B\|)^{1/2}}\right)\right\}.
			\end{split}
		\end{equation}
		Specifically, we will bound $\a\trans(\breve{\SSigma}^\ast - \hat\SSigma)\b$ and $\a\trans(\breve{\SSigma}^\ast - \SSigma)\b$ separately in the next two steps.
		\item We consider $\a\trans(\breve{\SSigma}^\ast - \hat\SSigma)\b$ first. Since
		\begin{equation*}
			\breve{\sigma}^\ast_{ij} - \hat{\sigma}^\ast_{ij} = \frac{1}{n_{ij}^\ast}\sum_{k=1}^n (X_{jk}\bar X_i^\ast + X_{ik}\bar X_j^\ast)S_{ik}S_{jk} - \bar X_i^\ast \bar X_j^\ast,
		\end{equation*} 
		$\a\trans(\breve \SSigma^\ast - \hat \SSigma^\ast )\b$ can be written as
		\begin{equation}\label{eq:a(breve - hat)b}
			\begin{split}
				& \a\trans(\breve\SSigma^\ast - \hat \SSigma^\ast )\b = \sum_{i, j=1}^p a_ib_j (\breve \sigma_{ij}^\ast - \hat \sigma_{ij}^\ast)\\
				= & \sum_{i, j = 1}^p a_i b_j \Bigg(\frac{\sum_{k=1}^n X_{ik}S_{ik}}{n_i^\ast}\cdot\frac{\sum_{l=1}^n X_{jl}S_{il}S_{jl}}{n_{ij}^\ast}\\
				& + \frac{\sum_{k=1}^n X_{ik}S_{ik}S_{jk}}{n_{ij}^\ast}\cdot\frac{\sum_{l=1}^n X_{jl}S_{jl}}{n_j^\ast} - \frac{\sum_{k=1}^n X_{ik}S_{ik}}{n_i^\ast}\cdot\frac{\sum_{l=1}^n X_{jl}S_{jl}}{n_j^\ast}\Bigg)\\
				= & \sum_{i,j = 1}^p\sum_{k, l = 1}^n X_{ik}X_{jl}a_ib_j \left(\frac{S_{ik}S_{il}S_{jl}}{n_i^\ast n_{ij}^\ast} + \frac{S_{ik}S_{jk}S_{jl}}{n_{ij}^\ast n_j^\ast} - \frac{S_{ik}S_{jl}}{n_i^\ast n_j^\ast}\right).
			\end{split}
		\end{equation}
		We can calculate from \eqref{eq:a(breve - hat)b} that 
		\begin{equation}\label{eq:Ea(breve - hat)b}
			\begin{split}
				& \left|\Ex\a\trans(\breve{\SSigma}^\ast - \hat{\SSigma}^\ast)\b\right|\\
				= & \left|\sum_{i,j = 1}^p\sum_{k = 1}^n \sigma_{ij}a_ib_j \left(\frac{S_{ik}S_{ik}S_{jk}}{n_i^\ast n_{ij}^\ast} + \frac{S_{ik}S_{jk}S_{jk}}{n_{ij}^\ast n_j^\ast} - \frac{S_{ik}S_{jk}}{n_i^\ast n_j^\ast}\right)\right|\\
				= & \left|\sum_{i, j=1}^p \sigma_{ij}\frac{a_i}{n_i^\ast}b_j + \sum_{i, j = 1}^p a_i\frac{b_j}{n_j^\ast} \sigma_{ij} - \sum_{k=1}^n\sum_{i, j=1}^p\frac{S_{ik}a_i}{n_i^\ast}\frac{S_{jk}b_j}{n_j^\ast}\sigma_{ij}\right|\\
				\leq & \left|\left(\frac{a_1}{n_1^\ast}, \ldots, \frac{a_p}{n_p^\ast}\right)\SSigma \b\right| + \left|\a\trans\SSigma \left(\frac{b_1}{n_1^\ast}, \ldots, \frac{b_p}{n_p^\ast}\right)^\top\right|\\
				& + \sum_{k=1}^n \left|\left(\frac{S_{1k}a_1}{n_1^\ast}, \ldots, \frac{S_{pk}a_p}{n_p^\ast}\right)\SSigma \left(\frac{S_{1k}b_1}{n_1^\ast}, \ldots, \frac{S_{pk}b_p}{n_p^\ast}\right)\trans\right| \\
				\leq & \|\SSigma_{A\times B}\|\frac{\|\a\|_2\|\b\|_2}{n_{\min}^\ast} + \|\SSigma_{A\times  B}\|\frac{\|\a\|_2\|\b\|_2}{n_{\min}^\ast}\\
				& + \sum_{k=1}^n\|\SSigma_{A\times B}\|\cdot \frac{1}{2}\left\{\left\|\left(\frac{S_{1k}a_1}{n_1^\ast}, \ldots, \frac{S_{pk}a_p}{n_p^\ast}\right)\right\|_2^2 + \left\|\left(\frac{S_{1k}b_1}{n_1^\ast}, \ldots, \frac{S_{pk}b_p}{n_p^\ast}\right) \right\|_2^2\right\}.
			\end{split}
		\end{equation}
		For the last term in \eqref{eq:Ea(breve - hat)b}, we have the following bound,
		\begin{equation*}
			\begin{split}
				& \sum_{k=1}^n\|\SSigma_{A\times B}\|\cdot \frac{1}{2}\left\{\left\|\left(\frac{S_{1k}a_1}{n_1^\ast}, \ldots, \frac{S_{pk}a_p}{n_p^\ast}\right)\right\|_2^2 + \left\|\left(\frac{S_{1k}b_1}{n_1^\ast}, \ldots, \frac{S_{pk}b_p}{n_p^\ast}\right) \right\|_2^2\right\}\\
				= & \|\SSigma_{A\times B}\| \sum_{k=1}^n\sum_{i=1}^p \frac{1}{2} \left(\frac{S_{ik}a_i^2}{n_i^{\ast 2}} + \frac{S_{ik}b_i^2}{n_i^{\ast 2}}\right)\\
				= & \|\SSigma_{A\times B}\| \sum_{i=1}^p \frac{1}{2}\left(\frac{a_i^2 + b_i^2}{n_i^\ast}\right)\\
				\leq & \|\SSigma_{A\times B}\| \sum_{i=1}^p \frac{a_i^2+b_i^2}{2n_{\min}^\ast} \leq \frac{(\|\SSigma_A\|\|\SSigma_B\|)^{1/2}}{n_{\min}^\ast}.
			\end{split}
		\end{equation*}
		Thus, by \eqref{eq:Ea(breve - hat)b} and the inequality above, we have
		\begin{equation}\label{ineq:Ea(breve - hat)b}
			\left|\Ex a^\top (\breve \SSigma^\ast - \hat{\SSigma}^\ast) b \right| \leq \frac{3(\|\SSigma_A\|\|\SSigma_B\|)^{1/2}}{n_{\min}^\ast}.
		\end{equation}
		The last term of \eqref{eq:a(breve - hat)b} can be treated as a quadratic form of the vectorization of $\X: \text{vec}(\X)\in \mathbb{R}^{pn}$. We note the last term as $\text{vec}(\X)\trans \Q \text{vec}(\X)$, where $\Q \in \mathbb{R}^{pn\times pn}$ and 
		$$\Q_{(i, k), (j, l)} = a_ib_j \left(\frac{S_{ik}S_{il}S_{jl}}{n_i^\ast n_{ij}^\ast} + \frac{S_{ik}S_{jk}S_{jl}}{n_{ij}^\ast n_j^\ast} - \frac{S_{ik}S_{jl}}{n_i^\ast n_j^\ast}\right),\quad 1\leq i, j\leq p, 1\leq k, l\leq n. $$
		$\Q$ has the following properties,
		\begin{equation}
			\begin{split}
				\|\Q\|_F^2 = & \sum_{i,j = 1}^p\sum_{k,l=1}^n a_i^2b_j^2 \left(\frac{S_{ik}S_{il}S_{jl}}{n_i^\ast n_{ij}^\ast} + \frac{S_{ik}S_{jk}S_{jl}}{n_{ij}^\ast n_j^\ast} - \frac{S_{ik}S_{jl}}{n_i^\ast n_j^\ast}\right)^2\\
				\leq & \sum_{i, j=1}^p a_i^2b_j^2\sum_{k, l =1}^n \left(2\frac{S_{ik}S_{il}S_{jl}}{n_i^{\ast 2}n_{ij}^{\ast 2}} + 2\frac{S_{ik}S_{jk}S_{jl}}{n_{ij}^{\ast 2}n_j^{\ast 2}} + \frac{S_{ik}S_{jl}}{n_i^{\ast 2}n_j^{\ast 2}}\right),\quad \text{since } S_{ik} \in \{0, 1\};\\
				\leq & \sum_{i,j=1}^p a_i^2b_j^2 \frac{5}{n_{\min}^{\ast 2}} = \frac{5\|\a\|_2^2\|\b\|_2^2}{n_{\min}^{\ast 2}} = \frac{5}{n_{\min}^{\ast 2}};
			\end{split}
		\end{equation}
		\begin{equation}
			\|\Q\| \leq \|\Q\|_F \leq \frac{\sqrt{5}\|\a\|_2\|\b\|_2}{n_{\min}^\ast} \leq \frac{\sqrt{5}}{n_{\min}^{\ast}}.
		\end{equation}
		For ${\rm vec}(\X) \in \mathbb{R}^{pn}$, since its segments $\{\X_{k}, k=1,\ldots, p\}$ are independent and $\X_k = \GGamma\Z_k$, we can further write $\text{vec}(\X) = \D_{\GGamma} \text{vec}(\Z)$, where $\D_{\GGamma}\in \mathbb{R}^{pn\times qn}$ is with $n$ diagonal blocks of $\GGamma$, $\text{vec}(\Z)$ is a $(qn)$-dimensional i.i.d. sub-Gaussian random vector. Based on Hanson-Wright's inequality (Theorem 1.1 in Rudelson and Vershynin \cite{Rudelson_Hanson_Wright}),
		\begin{equation}
			\begin{split}
				& \Pr\left\{\left|\a\trans\left(\breve{\SSigma}^\ast - \hat{\SSigma}^\ast\right)\b - \Ex\a\trans\left(\breve{\SSigma}^\ast - \hat{\SSigma}^\ast\right)\b\right|\geq x\right\}\\
				= & \Pr\left\{\left|\text{vec}(\X)\trans\Q \text{vec}(\X) - \Ex\text{vec}(\X)\trans\Q \text{vec}(\X)\right|\geq x\right\}\\
				= & \Pr\left[\left|\text{vec}(\Z)\trans\D_{\GGamma}\trans\Q\D_{\GGamma} \text{vec}(\Z) - \Ex\left\{\text{vec}(\Z)\trans\D_{\GGamma}\trans\Q\D_{\GGamma} \text{vec}(\Z)\right\}\right|\geq x\right]\\
				\leq & 2\exp\left\{-c\min\left(\frac{x^2}{\tau^4\|\D_{\GGamma}\trans\Q\D_{\GGamma}\|_F^2}, \frac{x}{\tau^2\|\D_{\GGamma}\Q\D_{\GGamma}\|} \right)\right\}.
			\end{split}
		\end{equation}
		Here $c>0$ is a uniform constant. Since $\Q$ is supported on $\{(i, k), (j, l): i\in A, j\in B\}$, we have $\D_{\GGamma}\trans\Q \D_{\GGamma} = \D_{\GGamma_A}\trans\Q_{A\times B}\D_{\GGamma_B}$. Here $\D_{\GGamma_A} \in \mathbb{R}^{|A|n\times qn}, \D_{\GGamma_B}\in \mathbb{R}^{|B|n\times qn}$ are with $n$ diagonal block $\GGamma_{A\times [q]}$ and $\GGamma_{B \times [q]}$, respectively, where $[q] = \{1, \ldots, q\}$. Since $\GGamma_{A\times [q]} \GGamma_{A\times [q]}\trans = \SSigma_{A}$, $\GGamma_{B\times [q]} \GGamma_{B\times [q]}\trans = \SSigma_{B}$, we know
		$$\|\D_{\GGamma_A}\| = \|\GGamma_{A\times [q]}\| \leq \|\SSigma_A\|^{1/2}, \quad \|\GGamma_{B\times [q]}\| \leq  \|\D_{\GGamma_B}\| \leq \|\SSigma_B\|^{1/2}. $$ 
		Then we further have
		\begin{equation}
			\begin{split}
				& \Pr\left\{\left|\a\trans\left(\breve{\SSigma}^\ast - \hat{\SSigma}^\ast\right)\b - \Ex\a\trans\left(\breve{\SSigma}^\ast - \hat{\SSigma}^\ast\right)\b\right|\geq x\right\}\\
				\leq & 2\exp\left\{-c\min\left(\frac{x^2}{\tau^4\|\D_{\GGamma_A}\trans\Q_{A\times B}\D_{\GGamma_B}\|_F^2}, \frac{x}{\tau^2\|\D_{\GGamma_A}\trans\Q_{A\times B}\D_{\GGamma_B}\|} \right)\right\}\\
				\leq & 2\exp\left\{-c\min\left(\frac{x^2}{\tau^4\|\D_{\GGamma_B}\|^2\|\D_{\GGamma_A}\trans\|^2\|\Q\|_F^2}, \frac{x}{\tau^2\|\D_{\GGamma_B}\|\|\D_{\GGamma_A}\trans\|\|\Q\|} \right)\right\}\\
				\leq & 2\exp\left[-c\min\left\{\frac{x^2}{\tau^4\|\SSigma_A\|\|\SSigma_B\|\|\Q\|_F^2}, \frac{x}{\tau^2(\|\SSigma_A\|\|\SSigma_B\|)^{1/2}\|\Q\|} \right\}\right]\\
				\leq & 2\exp\left[-c \min\left\{\frac{x^2n_{\min}^{\ast 2}}{\tau^4\|\SSigma_A\|\|\SSigma_B\| }, \frac{xn_{\min}^{\ast}}{\tau^2(\|\SSigma_A\|\|\SSigma_B\|)^{1/2}}\right\}\right].
			\end{split}
		\end{equation}
		We define $x' = \max \left\{x -3(\|\SSigma_A\|\|\SSigma_B\|)^{1/2}/n_{\min}^\ast , 0\right\}$, combining the inequality above and \eqref{ineq:Ea(breve - hat)b}, we have
		\begin{equation}\label{ineq:a(breve - hat)b}
			\begin{split}
				& \Pr\left\{\left|\a^\top\left(\breve \SSigma^\ast - \hat{\SSigma}^\ast\right)\b\right| \geq x\right\} \leq \Pr\left\{\left|\a^\top\left(\breve \SSigma^\ast - \hat{\SSigma}^\ast\right)\b - \Ex\a^\top\hat{\SSigma}^\ast \b\right| \geq x'\right\}\\
				\leq & 2\exp\left[-c \min\left\{\frac{(x')^2n_{\min}^{\ast 2}}{\tau^4\|\SSigma_A\|\|\SSigma_B\| }, \frac{x'n_{\min}^{\ast}}{\tau^2(\|\SSigma_A\|\|\SSigma_B\|)^{1/2}}\right\}\right]\\
				\leq & 2\exp\left[-c' \min\left\{\frac{x^2n_{\min}^{\ast 2}}{\tau^4\|\SSigma_A\|\|\SSigma_B\| }, \frac{xn_{\min}^{\ast}}{\tau^2(\|\SSigma_A\|\|\SSigma_B\|)^{1/2}}\right\} +C\max \left(\frac{1}{\tau^4}, \frac{1}{\tau^2}\right)\right]\\
				\leq & C\exp\left[-c' \min\left\{\frac{x^2n_{\min}^{\ast 2}}{\tau^4\|\SSigma_A\|\|\SSigma_B\| }, \frac{xn_{\min}^{\ast}}{\tau^2(\|\SSigma_A\|\|\SSigma_B\|)^{1/2}}\right\} \right].
			\end{split}
		\end{equation}
		In the last inequality above, we used a fact that $\tau$ is lower bounded by a uniform constant. This is due to Assumption \ref{as:sub-Gaussian} that $\Ex(Z) = 0$, $\var(Z) = 1$, $\Ex\exp(tZ)\leq \exp(t^2\tau^2/2)$. Then,
		$$\exp(4\tau^2/2) \geq\frac{1}{2}\left\{\Ex\exp(2Z) + \Ex\exp(-2Z)\right\} =\sum_{k=0}^\infty \frac{2^{2k}\Ex Z^{2k}}{(2k)!} \geq 2\Ex Z^2 = 2, $$
		which implies $\tau^2\geq \frac{1}{2}\ln (2)$.

		\item It is easy to see that $\Ex \breve{\SSigma}^\ast = \SSigma$, so $\Ex \a\trans(\breve\SSigma^\ast - \SSigma)\b = 0$. Then
		\begin{equation}\label{eq:a^T(breve-Sigma)b_1}
			\begin{split}
				& a\trans(\breve \SSigma^\ast - \SSigma)b\\
				= & \sum_{i,j= 1}^p a_i b_j \left(\frac{1}{n_{ij}^\ast}\sum_{k=1}^n X_{ik}X_{jk}S_{ik}S_{jk}\right) - \Ex \sum_{i,j= 1}^p a_i b_j \left(\frac{1}{n_{ij}^\ast}\sum_{k=1}^n X_{ik}X_{jk}S_{ik}S_{jk}\right)\\
				= & \sum_{k=1}^n \sum_{i, j=1}^p \left(\frac{a_ib_jS_{ik}S_{jk}}{n_{ij}^\ast} X_{ik}X_{jk} - \Ex  \frac{a_ib_jS_{ik}S_{jk}}{n_{ij}^\ast} X_{ik}X_{jk} \right)\\
				\triangleq & \sum_{k=1}^n \left(\X_k\trans \C^k \X_k - \Ex \X_k \trans\C^k \X_k\right)\\
				= & \sum_{k=1}^n \left(\Z_k\trans \GGamma\trans \C^k\GGamma \Z_k - \Ex \Z_k\trans \GGamma\trans \C^k\GGamma \Z_k\right).
			\end{split}
		\end{equation}
		Here $\C^k\in \mathbb{R}^{p\times p}$ is a matrix such that $C^k_{ij} = a_ib_jS_{ik}S_{jk}/n_{ij}^\ast$. Note that $\C^k$ is supported on $A\times B$, we can prove the following properties of $\C^k$.
		\begin{equation}
			\begin{split}
				\|\GGamma\trans\C^k\GGamma\|_F = & \sqrt{\tr\left(\C^k\GGamma\GGamma\trans\C^{k\top}\GGamma\GGamma\trans \right)}\\
				= & \sqrt{\tr\left(\C^k_{A\times B} \GGamma_{B\times [q]}\GGamma_{B\times [q]}\trans\C^{k\top}_{A\times B}\GGamma_{A\times [q]}\GGamma_{A\times [q]}\trans\right)} \\ 
				\leq & \|\GGamma_{B\times [q]}\| \|\GGamma_{A\times [q]}\| \sqrt{\tr(\C^k\C^{k\top})}\\
				\leq & (\|\SSigma_A\|\|\SSigma_B\|)^{1/2} \sqrt{\tr(\C^k\C^{k\top})};
			\end{split}
		\end{equation}
		\begin{equation}\label{ineq:GammaCGamma_fro_sum}
			\begin{split}
				& \sum_{k=1}^n \|\GGamma^\top \C^k \GGamma\|_F^2 \leq \|\SSigma_A\|\|\SSigma_B\| \|\C^k\|_F^2 =   \|\SSigma_A\|\|\SSigma_B\| \sum_{k=1}^n \sum_{i,j=1}^p \left(\frac{a_ib_jS_{ik}S_{jk}}{n_{ij}^\ast}\right)^2\\
				= & \|\SSigma_A\|\|\SSigma_B\| \sum_{k=1}^n \sum_{i, j=1}^p \frac{S_{ik}S_{jk}a_i^2b_j^2}{n_{ij}^{\ast 2}} = \|\SSigma_A\|\|\SSigma_B\| \sum_{i,j=1}^p \frac{a_i^2b_j^2}{n_{ij}^\ast} \\
				\leq & \|\SSigma_A\|\|\SSigma_B\| \frac{\|\a\|_2^2\|\b\|_2^2}{n_{\min}^\ast} = \frac{\|\SSigma_A\|\|\SSigma_B\|}{n_{\min}^\ast};
			\end{split}
		\end{equation}
		\begin{equation}\label{ineq:GammaCGamma_spe}
			\begin{split}
				\|\GGamma^\top\C^k\GGamma\| \leq  & \|\GGamma^\top\C^k\GGamma\|_F \leq (\|\SSigma_A\|\|\SSigma_B\|)^{1/2} \sqrt{\tr(\C^k\C^{k\top})}\\
				\leq & (\|\SSigma_A\|\|\SSigma_B\|)^{1/2} \sqrt{\sum_{i, j=1}^p \left(\frac{a_ib_j S_{ik}S_{jk}}{n_{ij}^\ast}\right)^2}\\
				\leq & (\|\SSigma_A\|\|\SSigma_B\|)^{1/2}\sqrt{\sum_{i,j=1}^p \frac{a_i^2b_j^2}{n_{\min}^{\ast 2}}}\leq \frac{\sqrt{\|\SSigma_A\|\|\SSigma_B\|}}{n_{\min}^\ast}.
			\end{split}
		\end{equation}
		Now, note that the last line of \eqref{eq:a^T(breve-Sigma)b_1} can be also equivalently written as
		$$\text{vec}(\Z)^\top \C^{con}\text{vec}(\Z)^\top - \Ex \text{vec}(\Z)^\top \C^{con}\text{vec}(\Z)^\top, $$
		$$\C^{con} =  \begin{bmatrix}
		\GGamma^\top \C^1 \GGamma & & \\
		& \ddots & \\
		& & \GGamma^\top \C^n \GGamma
		\end{bmatrix} \in \mathbb{R}^{(qn)\times (qn)},  $$
		where $\text{vec}(\Z)$ is the vectorization of $\Z$, which is an $qn$-dimensional i.i.d. sub-Gaussian vector. Based on the properties of $\C^k$ above, we have
		$$\|\C^{con} \|_F^2 = \sum_{k=1}^n \|\GGamma^\top \C^k \GGamma\|_F^2 \overset{\eqref{ineq:GammaCGamma_fro_sum}}{\leq} \frac{\|\SSigma_A\|\|\SSigma_B\|}{n_{\min}^\ast},$$
		$$\|\C^{con}\| \leq \max_{1\leq k\leq n} \|\GGamma^\top \C^k \GGamma\| \overset{\eqref{ineq:GammaCGamma_spe}}{\leq} \frac{(\|\SSigma_A\|\|\SSigma_B\|)^{1/2}}{n_{\min}^\ast}. $$
		Now applying Hanson-Wright's inequality (Theorem 1.1 in Rudelson and Vershynin \cite{Rudelson_Hanson_Wright}), we have
		\begin{equation}
			\begin{split}
				& \Pr\left\{\left|\text{vec}(\Z_k)\trans \C^{con}\text{vec}(\Z_k) - \Ex \text{vec}(\Z_k)\trans \C^{con}\text{vec}(\Z_k)\right|\geq x\right\}\\
				\leq & 2\exp\left\{-c \min\left(\frac{x^2}{\tau^4\|\C^{con}\|_F^2}, \frac{x}{\tau^2\|\C^{con}\|}\right)\right\} \\
				\leq & 2\exp\left\{-cn_{\min}^\ast\min\left(\frac{x^2}{\tau^4\|\SSigma_A\|\|\SSigma_B\|}, \frac{x}{\tau^2 (\|\SSigma_A\|\|\SSigma_B\|)^{1/2}}\right)\right\}.
			\end{split}
		\end{equation}
		Thus, 
		\begin{equation}\label{ineq:a(breve - Sigma)b}
			\begin{split}
				& \Pr\left\{\left|\a^\top (\breve \SSigma^\ast - \SSigma)\b\right|\geq x\right\}\\
				\leq & 2\exp\left\{-cn_{\min}^\ast\min\left(\frac{x^2}{\tau^4\|\SSigma_A\|\|\SSigma_B\|}, \frac{x}{\tau^2 (\|\SSigma_A\|\|\SSigma_B\|)^{1/2}}\right)\right\}.
			\end{split}
		\end{equation}
		Here $c$ is a uniform constant. Combining \eqref{ineq:a(breve - Sigma)b} and \eqref{ineq:a(breve - hat)b}, we have \eqref{ineq:a(hat - Sigma)b}.
		
		\item Next, we use the $\varepsilon$-net technique to give the bound on $\|\hat\SSigma^\ast_{A\times B} - \SSigma_{A\times B}\|$. Denote $\D^\ast = \hat{\SSigma}^\ast_{A\times B} - \SSigma_{A\times B}$. Suppose $S_{1/3}^{A}$ is the $(1/3)$-net for all unit vectors in $\mathbb{R}^{|A|}$; similarly $S_{1/3}^B$ is the $(1/3)$-net for all unit vectors in $\mathbb{R}^{|B|}$. Based on the proof of Lemma 3 in Cai et al. \cite{Cai_Zhang_Zhou_bandable}, we can let $\text{Card}(S_{1/3}^A) \leq 7^k$, $\text{Card}(S_{1/3}^B) \leq 7^k$. Since for all $\a, \a_0\in \mathbb{R}^{|A|}, \b,  \b_0\in \mathbb{R}^{|B|}$,
		\begin{equation}
			\begin{split}
				\left|\a\trans\D^\ast \b\right| - \left|\a_0\trans\D^\ast \b_0\right| \leq & \left|\a\trans\D^\ast \b - \a_0\trans\D^\ast \b_0\right| \leq \left|(\a - \a_0)\trans\D^\ast \b\right| + \left|\a_0\trans\D^\ast (\b - \b_0) \right|\\
				\leq & \left(\|\a- \a_0\|_2 + \|\b - \b_0\|_2\right)\|\D^\ast\|,
			\end{split}
		\end{equation}
		we have for all $\a\in \mathbb{R}^{|A|}, \b\in \mathbb{R}^{|B|}, \|\a\|_2 = \|\b\|_2 = 1$, we can find $\a_0 \in S_{1/3}^A, \b_0\in S_{1/3}^B$ such that $\|\a_0 - \a\|_2\leq 1/3, \|\b_0 - \b\|_2\leq 1/3$, then
		$$|\a^\top \D^\ast \b| \leq |\a_0^\top \D^\ast \b_0| + \frac{2}{3}\|\D^\ast\| \leq \sup_{\a_0\in S_{1/3}^A, \b_0\in S_{1/3}^B}|\a_0^\top \D^\ast \b_0| + \frac{2}{3}\|\D^\ast\|, $$
		$$\|\D^\ast\| = \sup_{\substack{\a\in \mathbb{R}^{|A|}, \b\in \mathbb{R}^{|B|},\\ \|\a\|_2 = \|\b\|_2 = 1}} |\a^\top \D^\ast \b| \leq \sup_{\a_0\in S_{1/3}^A, \b_0\in S_{1/3}^B}|\a_0^\top \D^\ast \b_0| + \frac{2}{3}\|\D^\ast\|, $$
		which yields
		\begin{equation}
			\|\hat{\SSigma}^\ast_{A\times B} - \SSigma_{A\times B}\| = \|\D^\ast\| \leq 3\sup_{\a_0\in S_{1/3}^A, \b_0\in S_{1/3}^B}|\a_0\trans\D^\ast \b_0|.
		\end{equation}
		Finally, by combining \eqref{ineq:a(hat - Sigma)b} and the inequality above, we know there exist uniform constants $C_1, c>0$ such that for all $t>0$,
		\begin{equation}
			\begin{split}
				& \Pr\left(\|\hat\SSigma^\ast_{A\times B} - \SSigma_{A\times B}\| \geq x\right) \leq  \Pr\left(\sup_{\a_0\in S_{1/3}^A, \b_0 \in S_{1/3}^B}|\a_0\trans\D^\ast \b_0| \geq \frac{x}{3} \right)\\
				\leq & C_1(7)^{|A| + |B|}\exp\left[-cn_{\min}^\ast \min\left\{\frac{x^2}{\tau^4\|\SSigma_A\|\|\SSigma_B\|}, \frac{x}{\tau^2(\|\SSigma_A\|\|\SSigma_B\|)^{1/2}}\right\}\right].
			\end{split}
		\end{equation}
		Since $|A|+|B|\leq 2|A\cup B|$, we have finished the proof of Lemma \ref{lm:covariance_spectral_b}. \quad $\square$
	\end{enumerate}
	
	{\noindent\bf Proof of Lemma \ref{lm:spectral_AB}.} Since $\SSigma$ is positive semi-definite, we can find the Cholesky decomposition such that $\SSigma = \V \V^\top$. Then $\SSigma_{A\times B} = \V_{A\times [p]}\V_{B\times [p]}^\top$ and
	\begin{equation*}
		\begin{split}
			\|\SSigma_{A\times B}\| = & \max_{\substack{\x\in \mathbb{R}^{|A|}, \y\in \mathbb{R}^{|B|}\\ \|\x\|_2 = \|\y\|_2 = 1}} \x^\top\V_{A\times [p]} \V_{B\times [p]}^\top \y \\
			\leq & \max_{\substack{\x\in \mathbb{R}^{|A|}, \y\in \mathbb{R}^{|B|}\\\|\x\|_2 = \|\y\|_2 = 1}} \left(\x^\top\V_{A\times [p]} \V_{A\times [p]}^\top\x \right)^{1/2} \left(\y^\top\V_{B\times [p]}\V_{B\times [p]}^\top \y\right)^{1/2}\\
			= & \max_{\substack{\x\in \mathbb{R}^{|A|}, \y\in \mathbb{R}^{|B|}\\\|\x\|_2 = \|\y\|_2 = 1}} \left(\x^\top\SSigma_{A}\x \right)^{1/2} \left(\y^\top\SSigma_B \y\right)^{1/2} = \|\SSigma_A\|\|\SSigma_B\|.
		\end{split}
	\end{equation*}
	Here we have used the Cauchy-Schwarz inequality. \quad $\square$
	
	\subsection{Proof of Theorem \ref{th:bandable}}
	
	Define $\B = (b_{ij})_{1\leq i,j\leq p}$ such that $b_{ij} = \sigma_{ij}$ if $i \in I_s$, $j\in I_{s'}$ and $|s - s'| \leq 1$, and 0 otherwise. Let $\bDelta = \SSigma - \B$. Then 
	$$\|\hat{\SSigma}^{\rm bt} - \SSigma\| \leq \|\hat{\SSigma}^{\rm bt} - \B\| + \|\bDelta\|. $$
	It is easy to see that
	$$\|\bDelta\| \leq \|\bDelta\|_{\ell_1} \leq \max_{i} \sum_{j: |i-j| \geq k}|\sigma_{ij}| \leq Mk^{-\alpha}. $$
	To bound $\|\hat{\SSigma}^{\rm bt} - \B\|$, note that
	$$\|\hat{\SSigma}^{\rm bt} - \B\| = \sup_{u\in \mathbb{R}^p: \|u\|_2 = 1} \left|\langle u, (\hat{\SSigma}^{\rm bt} - \B)u\rangle\right|. $$
	For any $u\in \mathbb{R}^p$, $\|u\|_2  =1$, we have
	\begin{equation*}
		\begin{split}
			\left|\langle u, (\hat{\SSigma}^{\rm bt} - \B)u \rangle\right| \leq & \sum_{s, s': |s - s'|\leq 1}\left|\left\langle u_{I_s}, (\hat{\SSigma}^\ast_{I_s\times I_{s'}} - \SSigma_{I_s\times I_{s'}})u_{I_{s'}} \right\rangle\right|\\
			\leq & \sum_{s, s': |s - s'|\leq 1} \|u_{I_s}\|_2\|u_{I_{s'}}\|_2 \|\hat{\SSigma}^\ast_{I_s\times I_{s'}} - \SSigma_{I_s\times I_{s'}}\|\\
			\leq & \left(\sum_{s, s': |s - s'|\leq 1} \|u_{I_s}\|_2\|u_{I_{s'}}\|_2\right) \left(\max_{|s - s'|\leq 1} \|\hat{\SSigma}^\ast_{I_s\times I_{s'}} - \SSigma_{I_s\times I_{s'}}\|\right).
		\end{split}
	\end{equation*}
	The Cauchy-Schwarz inequality yields
	\begin{equation}
		\sum_{s, s': |s - s'|\leq 1} \|u_{I_s}\|_2\|u_{I_{s'}}\|_2 \leq \frac{1}{2} \sum_{s, s': |s- s'| \leq 1} \left(\|u_{I_s}\|_2^2 + \|u_{I_{s'}}\|_2^2\right) \leq 3\sum_{s=1}^N \|u_{I_s}\|_2^2 = 3.
	\end{equation}
	Therefore, 
	\begin{equation*}
		\begin{split}
			\|\hat{\SSigma}^{\rm bt} - \SSigma\| \leq & \|\hat{\SSigma}^\ast - \B\| + \|\bDelta\|\leq   3 \max_{|s -  s'|\leq 1} \left\|\hat{\SSigma}^\ast_{I_s\times I_{s'}} - \SSigma_{I_s\times I_{s'}}\right\| + Mk^{-\alpha},
		\end{split}
	\end{equation*}
	which yields
	$$\Ex \|\hat{\SSigma}^{\rm bt} - \SSigma\|^2 \leq 18 \Ex \left(\max_{|s -  s'|\leq 1} \left\|\hat{\SSigma}^\ast_{I_s\times I_{s'}} - \SSigma_{I_s\times I_{s'}}\right\|\right)^2 + 2M^2k^{-2\alpha}. $$
	According to lemma \ref{lm:covariance_spectral_b}, there exists constant $C, c>0$ which only depend on $\tau$ such that for all $x>0$,
	\begin{equation}
		\begin{split}
			\Pr\left(\max_{|s - s'|\leq 1}\|\hat{\SSigma}_{I_s \times I_{s'}} - \SSigma_{I_s \times I_{s'}} \| \geq x\right)\leq & C\lceil \frac{p}{k}\rceil (49)^k \exp\left\{-c n_{\min}^\ast \min \left(\frac{x^2}{\|\SSigma\|^2}, \frac{x}{\|\SSigma\|}\right)\right\}.
		\end{split}
	\end{equation}
	Now we set $t = C'(k + \ln p)/n_{\min}^\ast$ for $C'$ large enough. The spectral norm risk satisfies
	\begin{equation}\label{ineq:ESigma^tb - Sigma}
		\begin{split}
			\Ex \|\hat{\SSigma}^{\rm bt} - \SSigma\|^2 \leq & 18\Ex \max_{|s - s'| \leq 1} \left\|\hat{\SSigma}^\ast_{I_s\times I_{s'}}\right\| + 2M^2k^{-2\alpha}\\
			\leq & 18\int_{0}^\infty \Pr\left(\max_{|s - s'|\leq 1}\|\hat{\SSigma}_{I_s\times I_{s'}} - \SSigma_{I_s\times I_{s'}} \|^2 \geq x\right)dx + 2M^2k^{-2\alpha}\\
			\leq & 18t + 18\int_t^\infty \Pr\left(\max_{|s - s'|\leq 1}\|\hat{\SSigma}_{I_s\times I_{s'}} - \SSigma_{I_s\times I_{s'}} \|^2 \geq x\right)dx + 2M^2k^{-2\alpha}\\
			\leq & 18t + C\lceil \frac{p}{k}\rceil (49)^k\int_{t}^\infty \exp\left\{-c'n_{\min}^\ast\min\left(x, x^{\frac{1}{2}}\right)\right\} dx + 2M^2k^{-2\alpha}\\
			\leq & 18t + C\lceil \frac{p}{k}\rceil (49)^k \frac{1}{n_{\min}^\ast}\exp\left(-c'n_{\min}^\ast t\right) + 2M^2k^{-2\alpha},
		\end{split}
	\end{equation}
	then \eqref{ineq:ESigma^tb - Sigma} yields
	\begin{equation}
		\Ex \|\hat\SSigma^{\rm bt} - \SSigma\|^2 \leq C\left(\frac{k+\ln p}{n_{\min}^\ast} + k^{-2\alpha}\right),
	\end{equation}
	where $C$ only depends on $\tau, M, M_0$. We can finally finish the proof of Theorem \ref{th:bandable} by taking $k = (n_{\min}^\ast)^{1/(2\alpha +1)}$.\quad $\square$

	\subsection*{Acknowledgments}
	
	We thank Tianxi Cai for the ovarian cancer data set and for helpful discussions. We also thank the Editor, the Associate editor, one referee and Zoe Russek for useful comments which have helped to improve the presentation of the paper.

\newpage
\subsection*{Appendix: Proofs}	
	
In this appendix we collect the proofs for the main results of the sparse covariance matrix estimation and Propositions \ref{pr:bandable_lower} and \ref{pr:sparse_lower}.
	
		\subsection*{Proof of Lemma \ref{lm:covariance}}
		
		The main strategy for the proof of this lemma is similar to that for  Lemma 2 in Cai and Liu \cite{Cai_Liu}. Without loss of generality, we can translate $X$ and assume that $\Ex \X = \boldsymbol{\mu} = 0$.
		First, we show the following property on $\theta_{ij}$
		\begin{equation}\label{ineq:theta_ij}
		c\sigma_{ii}\sigma_{jj} \leq \theta_{ij} \leq C\sigma_{ii}\sigma_{jj}.
		\end{equation}
		Here $c, C>0$ only depend on the distribution of $\Z$. Denote $\a, \b$ as the $i$-th and $j$-th row vector of $\GGamma$, then $\|\a\|_2^2 = \var(X_i) =  \sigma_{ii}$, $\|b\|_2^2 = \sigma_{jj}$. Recall that 
		$$\theta_{ij} = \var(X_iX_j - \sigma_{ij}) = \var(\a\trans\Z \b\trans\Z - \Ex \a\trans\Z \b\trans\Z),$$
		thus
		\begin{equation}
		\begin{split}
		\theta_{ij} =  \var\left(\Z\trans\a \b\trans\Z\right) \leq \Ex \left(\Z\trans\a \b\trans\Z\right)^2 \leq  \sqrt{\Ex (\Z\trans\a)^4 \Ex (\Z\trans\b)^4}.
		\end{split}
		\end{equation}
		Since 
		$$\Ex \left(\Z\trans\a\right)^4 \leq \left(\sum_{s = 1}^q a_s^4\Ex Z_s^4 + 6\sum_{1\leq s< t\leq q} a_s^2a_t^2 \Ex Z_s^2 \Ex Z_t^2\right) \leq 3\|\a\|_2^4\Ex Z^4 \leq C_\tau \sigma_{ii}^2,
		$$
		similarly $\Ex \left(\Z\trans\b\right)^4 \leq C_\tau \sigma_{jj}^2$, we know $\theta_{ij} \leq C_\tau \sigma_{ii}\sigma_{jj}$ for some constant $C_\tau$ only depending on $\tau$.
		
		On the other hand, since the entries of $\Z$ are i.i.d. with mean 0 and variance 1 and $\var(Z^2) > 0$, we know $\Ex Z^4 > (\Ex Z^2)^2 = 1$. We can calculate that
		\begin{equation*}
		\begin{split}
		\theta_{ij} = & \var(X_iX_j) = \Ex (X_iX_j)^2 - \left(\Ex X_iX_j\right)^2\\ 
		= & \Ex \left\{\sum_{s=1}^q a_sb_s Z_s^2 + \sum_{1\leq s<t\leq q}(a_sb_t + a_tb_s)Z_sZ_t\right\}^2 - \left(\sum_{s=1}^q a_sb_s\right)^2\\
		= & \sum_{s=1}^q a_s^2b_s^2 \Ex Z_s^4 + \sum_{1\leq s <t \leq q} \left(a_s^2b_t^2 + a_t^2 b_s^2 + 4a_sb_sa_tb_t\right)\Ex Z_s^2Z_t^2 - \left(\sum_{s=1}^q a_sb_s\right)^2\\
		= & \sum_{s=1}^q a_s^2b_s^2(\Ex Z^4 - 3) + \left(\sum_{s=1}^q a_s^2\right)\left(\sum_{t=1}^q b_t^2\right) + \left(\sum_{s=1}^q a_sb_s\right)^2.
		\end{split}
		\end{equation*}
		When $\Ex Z^4\geq 3$, it is clear that
		$$\theta_{ij} \geq \left(\sum_{s=1}^q a_s^2\right)\left(\sum_{s=1}^q b_s^2\right) = \|\a\|_2^2\|\b\|_2^2 = \sigma_{ii}\sigma_{jj};$$
		when $\Ex Z^4 < 3$, note $\xi = \Ex Z^4$, $x = \sum_{s: a_sb_s\geq 0} a_sb_s$, $y = -\sum_{s: a_sb_s<0}a_sb_s$, then $x, y\geq 0$ and
		\begin{equation*}
		\begin{split}
		\theta_{ij} \geq & -(3 - \xi)(x^2 + y^2) + (x-y)^2 + \frac{3 - \xi}{2} \left(\sum_{s=1}^q a_s^2\right)\left(\sum_{s=1}^q b_s^2\right) + \frac{\xi- 1}{2}\left(\sum_{s=1}^q a_s^2\right)\left(\sum_{s=1}^q b_s^2\right)\\
		\geq & -(3-\xi)(x^2 + y^2) + (x-y)^2 + \frac{3-\xi}{2}\left(\sum_{s=1}^q |a_sb_s|\right)^2 + \frac{\xi - 1}{2}\|\a\|_2^2\|\b\|_2^2\\
		= & -(3-\xi)(x^2 + y^2) + (x-y)^2 + \frac{3-\xi}{2}\left(x+y\right)^2 + \frac{\xi - 1}{2}\|\a\|_2^2\|\b\|_2^2\\
		\geq & \frac{\xi-1}{2}(x-y)^2 + \frac{\xi -1}{2}\sigma_{ii}\sigma_{jj}\geq c\sigma_{ii}\sigma_{jj}.
		\end{split}
		\end{equation*}
		Here $c = (\xi -1)/2$ only depends on the distribution of $Z$.
		
		Now we further normalize each row of $\GGamma$ such that $\|\GGamma_{i\cdot}\|_2= 1$ $\var(X_i) = \var(\GGamma_i\Z) = 1$ for $1\leq i\leq p$. The rest of the proof is essentially the same as Lemma 2 in Cai and Liu \cite{Cai_Liu} thus we will not go into details. Let 
		\begin{equation}
		\tilde{\theta}_{ij}^\ast = \frac{1}{n_{ij}^\ast} \sum_{k = 1}^n\left(X_{ik}X_{jk} - \tilde{\sigma}_{ij}^\ast\right)^2S_{ik}S_{jk}, \quad \tilde{\sigma}^\ast_{ij} = \frac{1}{n_{ij}^\ast}\sum_{k = 1}^n X_{ik}X_{jk}S_{ik}S_{jk}.
		\end{equation}
		We would like to show
		\begin{equation}\label{ineq:theta-tilde_theta}
		\Pr\left(\max_{ij} |\hat\theta_{ij}^\ast - \tilde{\theta}_{ij}^\ast| \geq C_1\sqrt{\ln p/n_{ij}^\ast}\right) = O(p^{-M}).
		\end{equation}
		Denote $X_i^{(j)\ast}$ as the average of $X_i$'s for those samples $X_i, X_k$ are both observed, i.e. $X_i^{(j)\ast} = \sum_{k=1}^n S_{ik}S_{jk} X_{ik}/n_{ij}^\ast$. Then,
		\begin{equation}\label{ineq:theta_hat_decompose}
		\begin{split}
		\hat{\theta}_{ij}^\ast = & \frac{1}{n_{ij}^\ast}\sum_{k = 1}^nS_{ik}S_{jk}\left(X_{ik}X_{jk} - \bar{X}_i^\ast X_{jk} - \bar X_j^\ast X_{ki} - \tilde{\sigma}_{ij}^\ast + \bar X_i^{(j)\ast} \bar X_j + \bar X_j^{(i)\ast} \bar X_i\right)^2\\
		= & \tilde{\theta}_{ij}^\ast + \frac{2}{n_{ij}^\ast}\sum_{k=1}^n S_{ik}S_{jk} \left(X_{ik}X_{jk} - \tilde{\sigma}_{ij}^\ast\right)\left(\bar X_i^{(j)\ast}\bar X_j^\ast + \bar X_j^{(i)\ast}\bar X_i^\ast - \bar X_i^\ast X_{jk} - \bar X_j^\ast X_{ik}\right)\\
		& + \frac{1}{n_{ij}^\ast}\sum_{k=1}^n S_{ik}S_{jk} \left(\bar X_i^{(j)\ast}\bar X_i^\ast + \bar X_j^{(i)\ast}\bar X_i^\ast - \bar X_i^\ast X_{jk} - \bar X_j^\ast X_{ik}\right)^2.
		\end{split}
		\end{equation}
		Similarly to Lemma 2 in Cai and Liu \cite{Cai_Liu}, we could have
		\begin{equation}
		\Pr\left(\max_{i,j} | \bar X_i^{(j)\ast}| \geq C_2\sqrt{\frac{\ln p}{n_{ij}^\ast}}\right) = O\left(p^{-M}\right),
		\end{equation}
		\begin{equation}
		\Pr\left(\max_{ij} \left|\frac{1}{n_{ij}^\ast}\sum_{k=1}^nS_{ik}S_{jk} X_{ik}^2 X_{jk}\bar X_j^\ast\right| \geq C_5\sqrt{\frac{\ln p}{n_{ij}^\ast}}\right).
		\end{equation}
		and similar bounds for the other terms in the right hand side of \eqref{ineq:theta_hat_decompose}. Hence we have proved \eqref{ineq:theta-tilde_theta}. By \eqref{ineq:theta_ij}, we can directly get 
		\begin{equation}\label{ineq:tilde_theta-theta}
		\Pr\left(\left|\tilde{\theta}_{ij}^\ast - \theta_{ij}\right|\geq \varepsilon\right) = O(p^{-M}),
		\end{equation}
		by applying the result in Lemma 2 in Cai and Liu \cite{Cai_Liu} on the samples $\X_{k}, k\in\{k: S_{ik} = S_{jk} = 1\}$. Combining \eqref{ineq:theta-tilde_theta} and \eqref{ineq:tilde_theta-theta}, we can proved \eqref{ineq:theta}.
	
		The proof of \eqref{ineq:sigma} is omitted here because it is essentially the same as that of Lemma 2 in \cite{Cai_Liu}.
		\quad $\square$
		
		\subsection*{Proof of Theorem \ref{th:sparse_cov}} 
		
		First without loss of generality, we can assume that $\boldsymbol{\mu} = \Ex \X_k =  0$. Based on Assumption \ref{as:sub-Gaussian}, we have for each $k$, $\X_k = \GGamma \Z_k$, where $\Z_k$ is an i.i.d. sub-Gaussian random vector. Based on the proof of Lemma \ref{lm:covariance}, we know $c\sigma_{ii}\sigma_{jj}\leq \theta_{ij} \leq C\sigma_{ii}\sigma_{jj}$, where $c, C$ are constants which only depend on the distribution of $\Z$. We will prove Theorem \ref{th:sparse_cov} in several steps.
		\begin{enumerate}
			\item For $\varepsilon>0$, we first consider the loss under the event that 
			\begin{equation}\label{ineq:situ_sparse_cov}
			Q = \left\{|\hat \sigma_{ij}^\ast - \sigma|/\hat{\theta}_{ij}^\ast \leq \delta \sqrt{\ln p/n_{ij}^\ast}, \forall 1\leq i, j\leq p, \quad \text{and}\quad \max_{ij}|\hat{\theta}_{ij}^\ast - \theta_{ij}|/(\sigma_{ii}\sigma_{jj}) \leq \varepsilon.\right\}.
			\end{equation}
			Since $|\hat{\sigma}_{ij}^\ast - \sigma_{ij}|\leq \delta \sqrt{\hat{\theta}_{ij}^\ast \ln p/n_{ij}^\ast} = \lambda_{ij}$, by Condition (1) of $T_{\lambda_{ij}}$, we have
			\begin{equation*}
			|T_{\lambda_{ij}}(\hat{\sigma}_{ij}^\ast) - \sigma_{ij}| \leq c_T|\sigma_{ij}|.
			\end{equation*}
			Besides, by condition (3) of $T_{\lambda_{ij}}$,
			\begin{equation*}
			\begin{split}
			& |T_{\lambda_{ij}}(\hat{\sigma}_{ij}^\ast) - \sigma_{ij}| \leq  |T_{\lambda_{ij}}(\hat{\sigma}_{ij}^\ast) - \hat{\sigma}_{ij}^\ast | + |\hat{\sigma}_{ij}^\ast - \sigma_{ij}| \overset{\eqref{ineq:situ_sparse_cov}}{\leq}  \lambda_{ij} + \delta \sqrt{\frac{\hat{\theta}_{ij}^\ast \ln p}{n_{ij}^\ast}}\\
			\leq & 2\delta \sqrt{\frac{\hat{\theta}_{ij}^\ast \ln p}{n_{ij}^\ast}}\leq 2\delta \sqrt{\frac{(\theta_{ij}+\varepsilon\sigma_{ii}\sigma_{jj})\ln p}{n_{ij}^\ast}}\leq  C\sqrt{\frac{\sigma_{ii}\sigma_{jj} \ln p}{n_{ij}^\ast}}.
			\end{split}
			\end{equation*}
			Since $n_{\min}^\ast \leq n_{ij}^\ast \leq n$, thus
			\begin{equation*}
			\begin{split}
			\|\hat{\SSigma}^{\rm at} - \SSigma\|_{\ell_1} \leq & \max_i \sum_{j=1}^p \left|T_{\lambda_{ij}}(\hat{\SSigma}_{ij}^\ast) - \sigma_{ij}\right|
			\leq \max_i \sum_{j=1}^p C\min\left\{|\sigma_{ij}|, \sqrt{\frac{\sigma_{ii}\sigma_{jj}\ln p}{n_{ij}^\ast}}\right\}\\
			\overset{\eqref{eq:Hc_np}}{\leq} & C c_{n, p} \sqrt{\frac{\ln p}{n_{\min}^\ast}} .
			\end{split}
			\end{equation*}
			Since $\hat{\SSigma}^{\rm at} - \SSigma$ is a symmetric matrix, we have $$\|\hat{\SSigma}^{\rm at} - \SSigma\|_{\ell_q} \leq \|\hat{\SSigma}^{\rm at} - \SSigma\|_{\ell_1} \leq Cc_{n, p}\sqrt{\frac{\ln p}{n_{\min}^\ast}}$$
			for all $1\leq q\leq \infty$.
			
			By Lemma \ref{lm:covariance}, we know \eqref{ineq:situ_sparse_cov} happens with probability at least $1 - O\left\{(\ln p)^{-1/2}p^{-\delta +2}\right\}$, which implies \eqref{ineq:sparse_cov_hp}.
			
			\item Next we consider \eqref{ineq:sparse_cov_E}. We apply Lemma \ref{lm:covariance_spectral_b} by restricting $\SSigma$ on $\{i, j\}\times \{i, j\}$ and set $A = \{i\}$, $B = \{j\}$, then there exists $C_1, c_1>0$ such that
			\begin{equation}\label{ineq:sparse_hat_sigma-sigma}
			\Pr\left(|\hat{\sigma}_{ij}^\ast - \sigma_{ij}| \leq x\right) \geq 1 - C_1\exp\left[-c_1 n_{ij}^\ast \min\left\{\frac{x^2}{\sigma_{ii}\sigma_{jj}} , \frac{x}{(\sigma_{ii}\sigma_{jj})^{1/2}}\right\}\right]
			\end{equation}
			holds for all $x>0$. Therefore,
			\begin{equation*}
			\Pr\left\{|\hat{\sigma}_{ij}^\ast - \sigma_{ij}| \leq x(\sigma_{ii}\sigma_{jj})^{1/2}, \forall 1\leq i, j\leq p \right\} \geq 1 - C_1p^2\exp\left\{-c_1n_{\min}^\ast\min(x, x^2)\right\}.
			\end{equation*}
			We also have
			$$|T_{\lambda_{ij}}(\hat{\sigma}^\ast_{ij}) - \sigma_{ij}| \leq c_T|\hat{\sigma}^\ast_{ij}| + |\sigma_{ij}| \leq (1+c_T)|\sigma_{ij}| + c_T|\hat{\sigma}^\ast_{ij} - \sigma_{ij}|. $$
			Thus,
			\begin{equation}\label{ineq:int_Qc_inter}
			\begin{split}
			& \Ex \|\hat{\SSigma}^{\rm at} - \SSigma\|^2_{\ell_1} = \int_Q \|\hat{\SSigma}^{\rm at} - \SSigma\|_{\ell_1}^2dP + \int_{Q^c} \|\hat{\SSigma}^{\rm at} - \SSigma\|_{\ell_1}^2dP\\
			\leq & Cc_{n, p}^2\frac{\ln p}{n_{\min}^\ast} + \int_{Q^c}\left(\max_{i} \sum_{j=1}^p |T_{\lambda_{ij}}(\hat{\sigma}_{ij}) - \sigma_{ij}|\right)^2dP\\
			\leq & Cc_{n, p}^2\frac{\ln p}{n_{\min}^\ast} + C\int_{Q^c} \left(\max_i \sum_{j=1}^p |\sigma_{ij}|\right)^2dP + C\int_{Q^c}\left(\max_i \sum_{j=1}^p |\hat{\sigma}_{ij}^\ast - \sigma_{ij}| \right)^2dP.\\
			\end{split}
			\end{equation}
			For the second term above, we have
			\begin{equation*}
			\begin{split}
			C\int_{Q^c} \left(\max_i \sum_{j=1}^p |\sigma_{ij}| \right)^2dP \leq & C\int_{Q^c} \left[\max_i \sum_{j=1}^p \min\left\{(\sigma_{ii}\sigma_{jj})^{1/2}, \frac{|\sigma_{ij}|}{\sqrt{\ln p/n_{\min}^\ast}}\right\}\right]^2dP\\
			\leq & C \Pr(Q^c)c_{n, p}^2 \leq C c_{n, p}^2 p^{-\delta+2} (\ln p)^{-1/2}.
			\end{split}
			\end{equation*}
			Based on the assumption that $p \geq (n_{\min}^\ast)^\xi$, $\delta \geq 4 + 1/\xi$, we have
			\begin{equation}\label{ineq:int_Qc_part_2}
			C\int_{Q^c} \left(\max_i \sum_{j=1}^p |\sigma_{ij}| \right)^2 \leq Cc_{n, p}^2\frac{\ln p}{n_{\min}^\ast}. 
			\end{equation}
			We denote $K = \max_i \sigma_{ii}$, then $K \leq c_{n, p}$. For the third term above in \eqref{ineq:int_Qc_inter}, we have
			\begin{equation*}
			\begin{split}
			& C\int_{Q^c}\left(\max_i \sum_{j=1}^p |\hat{\sigma}_{ij}^\ast - \sigma_{ij}| \right)^2dP \leq C p^2\int_{Q^c} \left(\max_{ij} |\hat{\sigma}_{ij}^\ast - \sigma_{ij}| \right)^2dP\\
			\leq & Cp^2 \int_{0}^\infty x \Pr\left(\{\max_{ij} |\hat{\sigma}_{ij}^\ast - \sigma_{ij}| \geq x\} \cap Q^c\right)dx\\ 
			= & Cp^2\int_0^{K} x \Pr\left(\{\max_{ij} |\hat{\sigma}_{ij}^\ast - \sigma_{ij}| \geq x\} \cap Q^c\right)dx\\
			& + Cp^2\int_K^{\infty} x \Pr\left(\{\max_{ij} |\hat{\sigma}_{ij}^\ast - \sigma_{ij}| \geq x\} \cap Q^c\right)dx\\
			\leq & Cp^2K^2\Pr(Q^c) + Cp^2 \int_{K}^\infty x\exp\left\{-c_1 n_{\min}^\ast \min\left(\frac{x^2}{\max_{i}\sigma_{ii}^2}, \frac{x}{\max_i \sigma_{ii}}\right) \right\}dx\\
			\leq & Cp^2c_{n, p}^2p^{-\delta+2}(\ln p)^{-1/2} + Cp^2 \int_K^\infty x\exp(-c_1 n_{\min}^\ast x/K)dx\\
			\leq & Cp^{-\delta+4}(\ln p)^{-1/2} c_{n, p}^2 + Cp^2c_{n, p}^2\exp(-c n_{\min}^\ast).
			\end{split}
			\end{equation*}
			Based on the assumption $\ln p = o((n_{\min}^\ast)^{1/3})$ and $p \geq (n_{\min}^\ast)^\xi$, $\delta \geq 4 + 1/\xi$, we have
			\begin{equation}\label{ineq:int_Qc_part3}
			C\int_{Q^c}\left(\max_i \sum_{j=1}^p |\hat{\sigma}_{ij}^\ast - \sigma_{ij}| \right)^2dP \leq Cc_{n, p}^2 \frac{\ln p}{n_{\min}^\ast}.
			\end{equation}
			Combining \eqref{ineq:int_Qc_part_2}, \eqref{ineq:int_Qc_part3} and \eqref{ineq:int_Qc_inter}, we have finished the proof of \eqref{ineq:sparse_cov_E} under the additional assumption that $p\geq (n_{\min}^\ast)^\xi$, $\delta \geq 4 + 1/\xi$.
			\quad $\square$
		\end{enumerate}
		
		\subsection*{Proof of Theorem \ref{th:support}.}
		By Lemma \ref{lm:covariance}, we know
		\begin{equation}\label{ineq:support_sigma}
		\Pr\left(|\hat{\sigma}_{ij}^\ast - \sigma_{ij}| \geq 2\sqrt{\frac{\ln p \hat{\theta}_{ij}^\ast}{n_{ij}^\ast}}, \exists 1\leq i, j\leq p\right) = O\left\{(\ln p)^{-1/2}\right\};
		\end{equation}
		for all $\varepsilon>0$,
		\begin{equation}\label{ineq:support_1}
		\Pr\left(|\hat{\theta}_{ij}^\ast - \theta_{ij}|/\sqrt{\sigma_{ii}\sigma_{jj}} \geq \varepsilon, \exists 1\leq i, j\leq p\right) = O(p^{-M}).
		\end{equation}
		Also by the proof of Lemma \ref{lm:covariance}, there exists $c>0$ such that 
		\begin{equation}\label{ineq:support_2}
		\theta_{ij}/\sqrt{\sigma_{ij}\sigma_{ij}} > c.
		\end{equation} 
		When $\delta=2$, the thresholding level is
		\begin{equation}\label{eq:support_lambda}
		\lambda_{ij} = 2\sqrt{\frac{\hat{\theta}_{ij}^\ast \ln p}{n_{ij}^\ast}}.
		\end{equation}
		Therefore,
		\begin{equation*}
		\begin{split}
		& \Pr\left\{\supp(\hat{\SSigma}^{\rm at}) \neq \supp(\SSigma)\right\}\\
		=& \Pr\left\{T_{\lambda_{ij}}(\hat{\sigma}^\ast_{ij}) = 0, \exists (i,j)\in \supp(\SSigma)\right\} + \Pr\left\{T_{\lambda_{ij}}(\hat{\sigma}^\ast_{ij}) \neq 0, \exists (i,j)\notin \supp(\SSigma)\right\} \\
		\leq & \Pr\left\{|\hat{\sigma}_{ij}^\ast| \leq \lambda_{ij}, \exists (i, j)\in \supp(\SSigma) \right\} + \Pr\left\{|\hat{\sigma}_{ij}^\ast| > \lambda_{ij}, \exists (i, j)\notin \supp(\SSigma) \right\}\\
		\overset{\eqref{eq:support_lambda}}{\leq} & \Pr\left(|\hat{\sigma}_{ij}^\ast - \sigma_{ij}| \geq 2\sqrt{\frac{\hat{\theta}_{ij}^\ast\ln p}{n_{ij}^\ast}}, \exists 1\leq i, j \leq p\right)\\
		& + \Pr\left\{|\sigma_{ij}| \leq 4\sqrt{\frac{\hat{\theta}_{ij}^\ast \ln p}{n_{ij}^\ast}},  \exists (i, j)\in \supp(\SSigma)\right\}\\
		\overset{\eqref{ineq:support_sigma}}{\leq} & O\left\{(\ln p)^{-1/2}\right\} + \Pr\left\{(4+\gamma)\sqrt{\frac{\theta_{ij}\ln p}{n_{ij}^\ast}} \leq 4\sqrt{\frac{\hat\theta_{ij}^\ast}{n_{ij}^\ast}}, \exists (i,j)\in \supp(\SSigma)\right\}\\
		\leq & O\left\{(\ln p)^{-1/2}\right\} + \Pr\left\{\left(\frac{4+\gamma}{4}\right)^2-1 \leq \frac{\hat{\theta}_{ij}^\ast - \theta_{ij}}{\theta_{ij}},\exists (i,j)\in \supp(\SSigma)\right\}\\
		\overset{\eqref{ineq:support_1}\eqref{ineq:support_2}}{=} & o(1),
		\end{split}
		\end{equation*}
		which means $\Pr\left\{\supp (\hat{\SSigma}^{\rm at}) \neq \supp(\SSigma)\right\} = o(1)$.
		\quad $\square$
		
		\subsection*{Proof of Propositions \ref{pr:bandable_lower} and \ref{pr:sparse_lower}.} 
		For given $n_0\geq 1$, we again consider a special pattern of missingness $\S_0$: 
			$$(\S_0)_{ij} = \left\{\begin{array}{ll}
			1, & 1\leq i\leq n_0, 1\leq j\leq p\\
			0, & n_0 +1\leq i\leq n, 1\leq j\leq p.
			\end{array}
			\right.$$
			Under this missingness pattern, $n_{\min}^\ast = n_0$, and the problem essentially becomes complete data problem with $n_0$ samples. Now, Propositions \ref{pr:bandable_lower}, \ref{pr:sparse_lower} directly follow Theorem 3 of Cai et al. \cite{Cai_Zhang_Zhou_bandable} and Theorem 2 of Cai and Zhou \cite{Cai_Zhou_covariance}, respectively.
	
\end{document}